\documentclass[11pt,a4paper]{article}
\usepackage{bm}
\usepackage{latexsym}
\usepackage{dcolumn}
\usepackage{amsmath,amsfonts,amssymb}
\usepackage{graphicx,epsfig}
\usepackage{psfrag}
\usepackage{color}

\usepackage{enumerate}
\usepackage[hang,flushmargin]{footmisc}
\usepackage{subcaption}
\usepackage{epstopdf} 
\usepackage{hyperref}

\usepackage{cite}

\usepackage{stackengine}
\usepackage{scalerel}

\newcommand{\be}{\begin{equation}}
\newcommand{\ee}{\end{equation}}
\newcommand{\bea}{\begin{eqnarray}}
\newcommand{\eea}{\end{eqnarray}}
\newcommand{\bse}{\begin{subequations}}
\newcommand{\ese}{\end{subequations}}
\newcommand{\bce}{\begin{center}}
\newcommand{\ece}{\end{center}}
\newcommand{\bfg}{\begin{figure}}
\newcommand{\efg}{\end{figure}}
\newcommand{\bit}{\begin{itemize}}
\newcommand{\eit}{\end{itemize}}
\newcommand{\bed}{\begin{description}}
\newcommand{\eed}{\end{description}}
\newcommand{\ben}{\begin{enumerate}}
\newcommand{\een}{\end{enumerate}}
\newcommand{\nn}{\nonumber}

\newcommand{\fr}{\frac}
\newcommand{\sq}{\sqrt}
\newcommand{\no}{\noindent}

\def\le {\left}
\def\ri {\right}
\def\a  {\alpha}

\def\c  {\gamma}

\def\d  {\delta}
\def\D  {\Delta}

\def\k  {\kappa}
\def\l  {\lambda}
\def\L  {\Lambda}
\def\m  {\mu}
\def\n  {\nu}
\def\o  {\omega}
\def\O  {\Omega}

\def\r  {\rho}
\def\th {\theta}

\def\s  {\sigma}

\def\vep {\varepsilon}

\newcommand{\cO}{\mathcal O}

\newcommand{\cR}{\mathcal R}

\newcommand{\ft}{\widetilde{f}}
\newcommand{\pt}{\widetilde{p}}

\newcommand{\rt}{\widetilde{\r}}

\newcommand{\vx}{\vec{\pmb x}}

\newcommand{\rmt}{\r^{(m)}}

\newcommand{\rbt}{\r^{(b)}}

\newcommand{\rct}{\r^{(c)}}

\newcommand{\rBt}{\r^{(B)}}

\newcommand{\UQ}{U_{_Q}}
\newcommand{\rx}{\r_{\!_X}}

\newcommand{\Omt}{\O^{(m)}}
\newcommand{\Obt}{\O^{(b)}}

\newcommand{\OLt}{\O^{(\L)}}

\newcommand{\Okt}{\O^{(k)}}

\newcommand{\pp}{p_{_0}}
\newcommand{\rmp}{\r^{(m)}_{_0}}
\newcommand{\Omp}{\O^{(m)}_{_0}}
\newcommand{\rbp}{\r^{(b)}_{_0}}
\newcommand{\Obp}{\O^{(b)}_{_0}}
\newcommand{\Okp}{\O^{(k)}_{_0}}
\newcommand{\rBp}{\r^{(B)}_{_0}}
\newcommand{\rcp}{\r^{(c)}_{_0}}

\newcommand{\OLp}{\O_{_0}^{(\L)}}

\newcommand{\OXp}{\O^{(X)}_{_0}}

\newcommand{\sw}{\mathsf w}

\newcommand{\wx}{\sw_{\!_X}}

\newcommand{\Rp}{R_{\!_0}}

\newcommand{\Hp}{H_{_{0}}}
\newcommand{\tp}{t_{_0}}

\newcommand*\rfra[2]{{}^{\scriptstyle{#1}}\!\!\diagup_{\!\!\scriptstyle{#2}}}

\begin{document}

\markboth{Saurya Das and Sourav Sur}
{A Unified Cosmological Dark Sector from a Bose-Einstein Condensate}

\title{\LARGE{A Unified Cosmological Dark Sector from a Bose-Einstein Condensate 
}}

\author{Saurya Das\footnote{Email: saurya.das@uleth.ca} \\ 
{\normalsize \em Theoretical Physics Group and Quantum 
Alberta, Department of Physics and Astronomy,}\\
{\normalsize \em University of Lethbridge, 4401 University 
Drive, Lethbridge, Alberta T1K 3M4, Canada}\\ \\
Sourav Sur\footnote{Corresponding Author. Email: sourav.sur@gmail.com, 
sourav@physics.du.ac.in} \\ 
{\normalsize \em Department of Physics and Astrophysics}\\
{\normalsize \em University of Delhi, New Delhi - 110007, 
India}}

\date{}
\maketitle


\begin{abstract}
We examine the viability of cosmological solution(s) describing a unified picture of the dark side of the universe from a Bose-Einstein condensate (BEC) of light bosons. 
The energy density of the BEC, together with its quantum potential, can indeed account for such a unification, in the sense that the (dust-like) cold dark matter and the dark energy components emerge from the same source.
In particular, the bulk of the dark energy can be attributed to the quantum potential, in the quantum corrected Raychaudhuri-Friedmann equation, when the `macroscopic' BEC wave-function is taken to be such that the corresponding probability density is construed as the energy density of the dusty fluid. 
However, there arises a purely quantum mechanical back-reaction, of even the visible baryons, on the effective dark energy and dark matter contents, which 
crucially determines the mass of the BEC. 
We determine the constraint on such a back-reaction, and hence on the BEC mass, from physical considerations, as well as estimate the same using recent observational data.

\end{abstract}

\vspace{10pt}
\no




\section{Introduction}

Despite numerous proposals and constructive efforts, a general consensus on the origin and structure of the cosmic {\em dark energy} (DE) and the {\em dark matter} (DM) remains elusive till date
\cite{CST-rev,FTH-rev,AT-book,wols-ed,MCGM-ed,BCNO-rev}.
While a certain degree of skepticism persists on the adequacy of General Relativity (GR) in this respect, wide attention is being drawn by the cosmological scenarios emerging not only from the classical formulations of alternative or modified gravity theories
\cite{NO-mgDE,tsuj-mgDE,CFPS-mg,JBE-mgDE,papa-ed,NOO-mg,ishak-mg},
but also from inherently quantum or semi-classical perspectives
\cite{suss-hol,DK-BECDE,hsu-holDE,li-holDE,CBJ-QCDE,MJ-pilDE-1,MJ-pilDE-2,nov-QG1,nov-QG2,nov-QG3,CL-QGDM,BTM-QCDE,singh-QCDE,AJM-QGDM,CF-QGDM}. 
Of course, the quantum effects on the standard cosmological picture are usually expected to have significance only in the very early universe, or in the spectrum of the density perturbations. However, certain approaches do point to their direct influence on the background (or homogeneous) level cosmic evolution even at the late-time regimes (red-shifts $z \lesssim 2$)
\cite{db1,db2,db3,DSS-QB}.

Observations not only favour the DE to be closely akin to a cosmological constant $\L$, but also indicate that the corresponding energy density is comparable to that of a mostly cold and non-relativistic DM, 
at (and near) the present epoch. As such, there arises the issue of {\em cosmic coincidence} with the ensuing idealized model, $\L$CDM, since the DE density (or $\L$) remains subdued, in comparison to the cold dark matter (CDM) density, for almost over the entire expansion history of the universe. Even the identification of $\L$CDM as being just a limiting case of a {\em dynamical} DE model, involving say, scalar field(s) such as quintessence or k-essence
\cite{CRS-quin,CLW-quin,tsuj-quin,AMS2000-kess,AMS2001-kess,MCLT-kess,schr-kess,SSSD-kess,SS-dquin},
is not enough to alleviate this problem, unless the DE and CDM components happen to be mutually interacting
\cite{amend-intDE,CPR-intDE,FP-intDE,BCHCSY-coupDE,diVal-coupDE,HVZ-coupDE,LLCST-coupDE}.
However, given the general characteristic difference of these components, it is often difficult to justify such interactions, unless they arise naturally, for e.g., via conformal transformations in certain scalar-tensor equivalent modified gravity formulations
\cite{amend-rev,FM-ST,frni-ST,BEPS-ST,TUMTY-ST,KT-ST,SSASB-MST,ASBSS-MSTda,MKSSS-MSTintDE,MKSSS-intDE}. 

A rather attractive proposition is of a {\em unified} cosmic dark sector, i.e. of the same fundamental origin of the DE, the CDM, and possible interaction(s) thereof. The redundancy of the coincidence problem is then obvious, however, with the DE and the CDM being reduced to mere structural artefacts of a {\em dark fluid}, within the standard paradigm of Friedmann-Robertson-Walker (FRW) cosmology. To be more precise, the unified dark sector may be thought of as being described by the dark fluid, composed of a pressure-free ({\em dust}-like) part, considered as the effective CDM, and a left-over, considered as the effective DE. If we assume, for simplicity, the energy-momentum conservation for the visible matter content of the universe, then so would be for the dark fluid, although the effective CDM and DE components of this fluid may be interacting. While such a system is shown to have a plausible realization from certain classical gravitational standpoints
\cite{BBM-uDE,BBPP-uDE,GNP-uDE,FFKB-uDE,CMV-mm,SVM-mm,CDS-MMT,SDC-MMT},
our objective in this paper is to see the implications of its inherently quantum (or semi-classical) description sought from the point of view of considering it to be due entirely to a {\em Bose-Einstein condensate} (BEC) of light bosons extending across cosmological length scales. Specifically, we intend to have a closer insight into the dark fluid picture that emerges from the BEC cosmological formalism developed in a series of earlier papers 
\cite{db1,db2,db3,DSS-QB},
by obtaining a solution of the corresponding evolution equations as general as possible, in an attempt to constrain the BEC mass parameter $m$, directly and robustly, using the observational data. 

The BEC cosmological formalism, which we consider here, essentially deals with a quantum correction to the Raychaudhuri-Friedmann equation, due to the bosons in the BEC being treated as quantum particles
\cite{sd},
which follow the Bohmian (quantal) trajectories, as a consequence of getting associated with a {\em quantum potential} $\UQ$
\cite{bohm-QT1,bohm-QT2,BHK-QT}.
Such a potential is derived from the amplitude of a generic wave-function $\Psi$, which may conveniently be chosen in a well-motivated Gaussian form (with a time-varying multiplicative factor), as in the earlier works
\cite{db1,db2,db3,DSS-QB}.
A large enough spread of the Gaussian (close to $\Hp^{-1}$, where $\Hp$ is the Hubble's constant) then implies that $\Psi$ is a `macroscopic' wave-function describing the BEC, which extends over such a large length scale. Together with the corresponding probability density $|\Psi|^2$, perceived as the BEC energy density $\rBt$ in a semi-classical approach, the quantum potential $\UQ$ consequently describes the evolution of the dark fluid.

Note that there is {\it a priori} no reason to assume the energy density and pressure of the dark fluid to be due entirely to $\rBt$ and $\UQ$ respectively. In fact, the BEC density $\rBt$ cannot account for the total effective CDM density, as the entire bulk of physical constituents of the universe (including even the visible baryons) back-reacts on the metric structure of space-time, by virtue of the quantum corrected Raychaudhuri-Friedmann equation. We have demonstrated this in a preceding work
\cite{DSS-QB},
by referring specifically to $\L$CDM, which is shown to be an exact solution of the BEC cosmological equations. It has also been argued therein, from certain physical considerations, that such a {\em quantum back-reaction} (QB) is very {\it mild}, and more interestingly, its magnitude (measured by some parameter $\vep$) is roughly proportional to the inverse of the BEC mass $m$. Therefore, the lower the value of $\vep$, the more is the enhancement of $m$ from the corresponding Hubble value $m_{_H} \simeq 10^{-32}\,$eV. However, $\vep$ cannot vanish altogether, because then $m$ would blow up. Thus, no matter how mild the QB is, it can always be regarded as an effect of significance from the point of view of obtaining a lower bound on the mass ratio $m/m_{_H}$. For instance, considering $\rBp < \rmp$, i.e. the BEC density is less than the total effective matter density at the present epoch $t=\tp$, it has been shown in ref.
\cite{DSS-QB}
that $\vep < 0.136$, using the Planck 2018 best fit cosmological parametric estimations for $\L$CDM
\cite{Planck18-CP}.
This in turn implies $m > 2.182 \, m_{_H}$, if one uses the corresponding best fit value of the DE density parameter at $\tp$, viz. $\OLp = 0.6889$.

Nevertheless, such a bound seems too loose to imply anything conclusively. For instance, it is loose enough compared to the typical mass limit $m \gtrsim 10^{-24}\,$eV predicted observationally, for suppressing small scale structures of commonly known ultralight bosonic DM candidates, such as {\em axions}
\cite{MH-UBDM,fer-UBDM,marsh-UBDM,MN-UBDM,APYMB-UBDM,CAMD-UBDM,HOTW-UBDM,HGMF-UBDM}.
Now, $m \gtrsim 10^{-24}\,$eV corresponds to $\vep \lesssim 10^{-8}$, meaning that the QB is negligible enough to leave the BEC density accounting for the effective CDM density in entirety. However, such a comparison (of the bounds) may not be quite justified here, since the scenario being still a unified one (regardless of the value of $\vep$), the BEC cannot be responsible for only the CDM, unlike what considered mostly in the literature 
\cite{HBG-FDM,wang-BEC_DM,FM-BEC_DM,FMT-BEC_DM,BH-BEC_DM,lopez-BEC_DM,sik-BEC_DM,chav-BEC_DM,harko-BEC_DM,DG-BEC_DM,KL-BEC_DM,HM-BEC_DM,SRM-BEC_DM,BCL-BEC_DM,LRS-BEC_DM,li-BEC_DM,EMM-BEC_DM,DKG-BEC_DM,giel-BEC_DM,david-BEC_DM,DKKG-BEC_DM,SCB-BEC_DM}
(see also
\cite{casadio1,casadio2}).
%
%
Besides, one cannot infer with any conviction that an enormously weak QB, or(and) the $\L$CDM solution is(are) the only one(s) plausible in this formalism. After all, note that the bound $\vep < 0.136$, obtained from just a physical consideration, and using merely the Planck 2018 best fit parametric estimates for $\L$CDM, is so loose that it even allows a BEC mass as low as $10^{-31}\,$eV or so, which typically corresponds to that of a scalar field DE candidate. 
%
%
%
%
%
%
%
%
It is therefore imperative to look for a more robust and tighter bound on $\vep$, by estimating this parameter directly from the observational results, and that too for a solution of the BEC cosmological equations as general as possible. 

At least, a solution more general than $\L$CDM is required, in order that the corresponding expression for the Hubble parameter $H(z)$ explicitly involves the parameter $\vep$, whose statistical estimation can therefore be made, using the commonly considered Pantheon Supernovae type-Ia (SN-Ia) and the observational Hubble data
%
%
\cite{scol,panth,MPCJM-OHD,YRW-OHD,RDR-OHD,mors}.
The determination of such a solution is nonetheless a hard proposition, given the highly complicated form of the quantum potential $\UQ$ thus encountered. 
%
However, the strong observational concordance on $\L$CDM makes one expect that any deviation from it, which implies a dynamical evolution of the effective DE component, is mild enough, in the sense that the cosmological parametric estimates do not breach the corresponding error margins for $\L$CDM up to $1\s$ or $2\s$ at most.
It therefore suffices to attempt for a solution which may be an approximated one, but generalizes the $\L$CDM solution of the BEC cosmological equations to a certain extent. In fact, one may exploit the smallness of $\vep$ to obtain an approximated solution more general than $\L$CDM, by resorting to a rather simplified expression for $\UQ$ that retains terms up to say, linear in $\vep$. Such a solution would, by construction, imply $H(z)$ explicitly involving $\vep$, and hence the latter can be constrained directly from the Pantheon and the Hubble observations.


Let us briefly outline the organization of this paper. We begin the section\,\ref{sec:BEC-Cosm} comtemplating on a general cosmological formulation with a BEC of ultralight bosons in the standard spatially flat FRW framework. Supposing the BEC to extend over the cosmological length scale, we look for a plausible semi-classical picture in subsection \ref{sec:DFdescrip}, with a prior objective of examining 
the natural emergence of a unified cosmic dark sector. While the latter can conveniently be described by an evolving dark fluid, mentioned above, we endeavour to pinpoint the criteria and assertions that are essential for its realization, in subsection \ref{sec:DFcriteria}. As we are primarily interested in studying the late-time cosmic evolution in this paper, we assume (for brevity) the visible matter content of the universe to be entirely in the form of a {\em dust} (of baryons). Accordingly, our course in section \ref{sec:BEC-DS} is set as follows:
\ben[(a)]
\item
working out the corresponding evolution equations in subsection 
\ref{sec:BEC-Cosm-eqs},
\item 
illustrating the features of the exact $\L$CDM solution in subsection\,\ref{sec:BEC-LCDM},
\item 
realizing the quantum back-reaction (QB) and its significance in subsection\,\ref{sec:BEC-QB}, and 
%
\item 
deriving a rather general solution analytically in subsection \ref{sec:BEC-dynDE}, by approximating the quantum potential $\UQ$ up to the leading order in the QB parameter $\vep$. 
\een 
While such a solution makes it suitable for us to carry out the statistical parametric estimation using the SN-Ia and the observational $H(z)$ data, we have the objective of figuring out, in section \ref{sec:BEC-est}, the extent to which the scenario deviates from $\L$CDM and the BEC mass $m$ gets enhanced from its Hubble value $m_{_H} \simeq 10^{-32}\,$eV, up to $1\s$. 
%
Nevertheless, there remains the scope to look for additional effects when the assumption of the spatial flatness of the universe is dropped. Examining whether such effects are of significance is our point of study in section \ref{sec:BEC-kDS}, wherein we look to generalize the cosmological solution obtained in subsection \ref{sec:BEC-dynDE} in presence of a non-vanishing curvature constant $k$. Finally, we summarize the course of the entire formulation, and the results of the subsequent analysis, highlighting the key aspects thereof, while concluding in section\,\ref{sec:concl}.

\bigskip
\no 
{\sl Convention and notation}: Everywhere, we use metric signature $\, (-,+,+,+)$, natural units (with the speed of light $c = 1$), and denote the metric determinant by $g$, the gravitational coupling factor by $\, \k = \sq{8 \pi G}$ (where $G$ is the Newton's constant), and the values of all quantities at the present epoch by a subscript $0$.

\section{Cosmology with a BEC: the General Formalism} \label{sec:BEC-Cosm}

Let us resort to the formal aspects of the standard (homogeneous and 
isotropic) FRW cosmology, by first referring to the usual perfect fluid 
description, in terms of the fluid velocity $u^\a$, the energy density 
$\rho$ and pressure $p$. In principle, this may be realized by a 
collection of perfect fluids as well, in which case their individual 
energy densities and pressures would sum up to give $\rho$ and $p$ 
respectively. The classical general relativistic principles imply that 
the cosmic evolution is governed by the Friedmann equations, which, 
under the assumption of spatial flatness, read as
\bea 
&& H^2(t) =\, \fr{\k^2} 3 \, \rho(t) \,\,,
\label{cosm-eq1}\\
&& \dot H(t) +\, H^2(t) =\, -\, \fr{\k^2} 6 \Big[\r(t) +\, 3 p(t)\Big] \,,
\label{cosm-eq2}
\eea
where $\k^2 = 8\pi G$, and $H(t) = \dot a(t)/a(t)$ is the Hubble 
parameter, with $a(t)$ being the FRW scale factor, and the overhead 
dot $\{\cdot\}$ denoting derivative with respect to the cosmic time $t$. 

Nevertheless, what would happen when a constituent fluid is inherently 
quantum in nature, say for e.g. a condensate of bosonic particles (i.e. 
a BEC) of rest mass $m$? The answer to this could be revealed from the 
explicit computation of the quantum corrections to the Raychaudhuri 
equation carried out by one of us (SD) in an earlier work 
\cite{sd}, 
by considering the Bohmian (quantal) particle trajectories
\cite{bohm-QT1,bohm-QT2,BHK-QT},
which are relevant in this case, in lieu of classical geodesics, as the 
bosons in the BEC are themselves quantum particles. 
In particular, taking the `quantum' fluid (or the BEC) wave-function in 
the generic form 
%
%
\be \label{WF}
\Psi (\vx,t) =\, \cR (\vx,t) \, e^{i \, S (\vx,t)} \,, 
\qquad \mbox{with} \quad
\cR, S \in \mathbb{R}\,\,,
\ee
the modified (or, quantum corrected) Raychaudhuri equation, describing 
the flow of the quantal trajectories parametrized by an affine 
parameter $\l$, is shown to be 
%
%
\be \label{QRE}
\fr{d\th}{d\l} =\, -\, \fr{\th^2} 3 -\, R_{\m\n} u^\m u^\n +\, 
\fr{\hbar^2}{m^2} \, h^{\m\n} 
\le(\fr{\square \cR} \cR\ri)_{\!;\m\,;\n} \,\,,
\ee
%
where $R_{\m\n}$ is the Ricci tensor, $\square \cR \equiv 
\cR^{;\a}_{~;\a}\,$, and $\th = h^{\m\n} u_{\m;\n}$ is the expansion 
scalar, with $\,h_{\m\n} = g_{\m\n} + u_\m u_\n\,$ denoting the 
induced metric (or the projection tensor onto the hyperplanes 
orthogonal to $u^\m$). Of course, the shear and vorticity terms are 
ignored here, as they have no relevance to the homogeneous-isotropic
cosmology. Hence, with the usual identifications 
\be \label{expansion}
\th =\, 3 H \quad \mbox{and} \quad R_{\m\n} u^\m u^\n =\, 
\frac{\k^2} 6 \big(\rho +\, p\big) \,,
\ee
the above equation (\ref{QRE}) reduces to 
\be \label{QRE0}
\fr{\ddot a} a \equiv\, \dot H +\, H^2 =\, -\, \fr{\k^2} 6 
\Big(\rho +\, 3 p\Big) +\, \fr 1 3 \, \UQ \,\,,
\ee
where the quantity 
%
\be \label{QP}
\UQ =\, \fr{\hbar^2}{m^2} \, h^{\m\n} \,  \le(\fr{\square \cR} 
\cR\ri)_{\!;\m\,;\n} \,\,,
\ee 
is interpreted as a {\em quantum potential} 
\cite{bohm-QT1,bohm-QT2,BHK-QT,sd}.

\subsection{The semi-classical dark fluid description} 
\label{sec:DFdescrip}

It is natural to identify Eq.\,(\ref{QRE0}) as the semi-classical version of the second order Friedmann equation (\ref{cosm-eq2}), simply under the replacement 
$\, p \rightarrow p - 2 \UQ/(3\k^2)$.
However, one needs to be cautious in making physical interpretations out of such a replacement. In fact, marking the 
$\UQ$-term
solely as the effective pressure of the quantum fluid is not quite tenable, 
because there is in principle nothing that can prevent the potential $\UQ$ to make a contribution to the system's total energy density $\rho$ as well. After all, even in a semi-classical approach, the compatibility of the Raychaudhuri equation with the Einstein's equations implies that the conservation relation 
\be \label{consv-eq}
\dot \rho =\, -\, 3 \, H \le(\rho +\, p\ri) \,,
\ee 
is not independent of the cosmic evolution equations, which must therefore be the Friedmann equations (\ref{cosm-eq1}) and (\ref{cosm-eq2}). In other words, one may reckon the semi-classical cosmic evolution to be described by Eqs.\,(\ref{cosm-eq1}) and (\ref{cosm-eq2}), despite the following:
\bit 
\item The total energy density $\rho$ is not simply given by that of the classical fluid matter content (the visible baryons, background radiation, etc.) plus that of the BEC (perceived semi-classically as the corresponding probability density $|\Psi|^2 = |\cR|^2$).
\item The total pressure $p$ is not just the classical fluid pressure (if any) 
augmented with $\, - 2 \UQ/(3\k^2)$.
%
\eit 
As such, a {\em unified} cosmic dark sector can emerge naturally, 
since by all means, the (quantum) fluid description of the BEC and 
the associated quantum potential $\UQ$ can determine the evolution 
profile of a {\em dark fluid}, that supposedly encodes the 
effective DE and DM components of the universe, and interaction(s) 
thereof. In other words, the dark fluid can be thought of as 
having energy density given by $\rho$ minus the classical fluid 
density, and pressure given by $p$ minus the classical fluid 
pressure (if any). Therefore, if the classical fluid obeys the 
energy-momentum conservation law, so would the dark fluid, 
regardless of its constituents -- the effective DE and DM -- 
not being conserved.

\subsection{Criteria and assertions for a consistent picture} 
\label{sec:DFcriteria}

Let us enlist the criteria, and subsequent assertions for a 
concrete BEC cosmological formulation, particularly from the 
point of view of the dark fluid description:
\bit
\item The foremost criterion is, of course, the large scale isotropy 
of the universe, which requires the embedded BEC wave-function to be 
spherically symmetric, i.e. $\Psi(\vx,t) \equiv \Psi (r,t) \,$, where 
the variable $r = \sq{h_{ij} x^i x^j}$ denotes the proper radial 
coordinate distance (which is implicitly time-dependent\footnote{
Note that $h_{ij} = h_{ij} (t) = a^2(t) \d_{ij}$. So, $\, r = r(t) = x \, a(t)$, where $x = \sq{\d_{ij} x^i x^j}$ is the comoving radial coordinate. 
}). Accordingly, the BEC density would in general be given by $|\cR (r,t)|^2$, where $\cR(r,t)$ is the radial part of $\Psi(r,t)$. Nevertheless, in limiting one's attention to the background (homogeneous) cosmological level in this paper, it suffices to resort to the purely time-dependent parts of the generic ($x,t$-dependent) BEC density and the associated quantum potential, denoting  them respectively as $\rBt (t)$ and $\UQ (t)$ henceforth. 
%
The remaining (inhomogeneous) parts can be treated as perturbations over the of the background level configuration determined by solving the Friedmann equations, which is commonly done in cosmological studies in a given setup. In other words, those inhomogeneous parts can be thought of as contributing to the total density and pressure perturbations in the effective dark fluid picture, in the usual sense of studying the metric perturbations for the entire system. However, we are leaving this for a future work, and focus only on the background cosmological evolution herein.
%
\item 
Next comes the question of suggesting the form of $\Psi(r,t)$. For this, it is worth resorting to the limiting Newtonian cosmological picture, given that the persistence of the BEC depends entirely on how small its mass $m$ is, so that the critical temperature $T_c\,$, below which the BEC is formed, exceeds the rapidly decreasing ambient temperature of the universe, as the latter evolves, at all epochs subsequent to the BEC formation (see the fourth point below for a further clarification). This, in turn, implies that well within the Hubble radius $\Hp^{-1}$, the bosons in the BEC are slow enough to be treated almost as non-relativistic
\cite{db1,db2}.
Now, for a mass $m$ placed on the surface of a homogeneous sphere of radius $r$ and density $\rBt$, we have the Newtonian gravitational equation
\be \label{Newt-eq} 
m \, \ddot r(t) =\, -\, \fr{G\, m}{r^2(t)} \le[\fr{4 \pi} 3 \, r^3 (t) \, \rBt(t)\ri] \,,
\ee 
which can be recast in the form of that of a harmonic oscillator, 
\be \label{Newt-eq1}
\ddot r (t) +\, \o^2 (t) \, r(t) =\, 0 \,\,,
\ee
albeit with a time-varying frequency 
\be 
\o(t) =\, \fr{4 \pi \,G} 3 \, \rBt(t) \,\,.
\ee 
Nevertheless, by the above argument, the time-dependence of the frequency can be considered as weak enough at length scales much smaller than $\Hp^{-1}$, wherein the Newtonian limit is fairly tenable. Therefore, in the usual quantum mechanical description, it is reasonable to suggest the BEC wave-function as that of the standard stationary state harmonic oscillator, module a time-dependent factor that can take care of the overall evolution of the BEC density $\rBt(t)$ at larger (cosmological) scales. In particular, going with the common supposition in the literature, that all the bosons in the BEC are in their ground state (of energy $E_0$), the harmonic oscillator wave amplitude being a Gaussian, we have the following suggested form of the BEC wave-function
\cite{db1,db2,db3,DSS-QB}
(see also
\cite{DS-EG1,DS-EG2,DS-EG3,DS-EG4}):
\be \label{WF1}
\Psi (r,t) =\, R(t) \, e^{-r^2/\s^2} \, e^{- i E_0 t/\hbar} \,, 
\ee
where $R(t)$ denotes the purely temporal modulating factor, and $\s$ is a real and positive constant parameter that denotes the spread of the Gaussian.
\item 
Now, from the semi-classical perspective, the purely temporal part of the probability density, viz. $|R(t)|^2$, being reckoned as the BEC density $\rBt(t)$ at the background (homogeneous) cosmological level, one has the obvious criterion of having the latter emulating the energy density of a dust-like fluid, or the CDM:
\be \label{rhoB} 
\rBt (t) =\, \rBp \, a^{-3} (t) \,, \quad \mbox{where} \quad 
\rBp \equiv \rBt(\tp)\,.
\ee
This necessitates the stipulation\footnote{
Note that a stipulation of the modulating factor $R(t)$ is always legitimate, once we consider normalizability of the BEC wave-function (\ref{WF1}) individually over each time slice, i.e.
\be \label{WF-norm}
\int d^3 x \sq{h(t)} \, \big\vert\Psi(r,t)\big\vert^2
\bigg\vert_{t \,=\, \mbox{\scriptsize constant}}
 =\, 1 \,,
\ee
where $h (t)$ denotes the corresponding metric determinant.
} 
\be \label{Rt}
R(t) =\, R_{_0} \, a^{- 3/2} (t) \,, \quad \mbox{where} \quad R_{_0} 
\equiv R(\tp) = \le[\rBp\ri]^{1/2}\,.
\ee 
%
%
%

%
%
%
\item 
Consequently, it rolls back to the question of asserting whether the BEC can actually account for the CDM, i.e. whether such a proposition is at all conducive to a cosmic evolution consistent with the observations. Indeed, as shown in ref. 
\cite{db1},
the BEC formation in the very early universe is plausible for an ideal gas of neutral bosons, provided they are ultralight. The precise mass bound is $m \lesssim 6\,$eV, which is obtained by demanding that the critical temperature $T_c\,$ (below which the BEC is formed) must exceed the ambient temperature of the universe, $T = 2.7 a^{-1}$ (in $K$), at all epochs (see also 
\cite{db2}
and the references therein). 
%
%
Specifically, $T_c$ is shown to vary as $m^{-1/3} a^{-1}\,$, when all the bosons in the BEC are in their ground state
\cite{db1}.
So, the smaller the value of $m$ the smaller is the ratio $T/T_c$, irrespective of the rapid decrease of $T$ with the expansion of the universe. This implies that for a small enough $m$, the BEC can form at a very low value of the scale factor $a$, whence $T$ is very high, and once formed, the BEC would persist forever, since $T/T_c$ is independent of $a$. Moreover, the ratio of the number of bosons within the BEC to the total number in a certain volume being given by $\, 1 - (T/T_c)^3 \,$, one infers that the BEC can eventually subsume most of the available light bosons in the universe, and thereby match up with the observed CDM content at the present epoch, at least to the correct order of magnitude 
\cite{db1,db2,db3}.
The BEC density $\rBt$ cannot, however, account for the entire CDM density of the universe at any epoch, because of a quantum {\em back-reaction}
\cite{DSS-QB}, 
whose significance would be discussed in due course.
%
\item Finally, it remains to be sorted out the conditions for the 
viability of the emerging scenario, which makes it imperative to 
obtain not only a suitable solution of the cosmic evolution 
equations, but also the estimates or bounds on the relevant 
parameters, such as the BEC mass $m$, using the observational data 
or results. This is the main purpose of this work, and it is here 
the quantum back-reaction (mentioned above) plays the utmost crucial 
role, demonstrated in what follows.
\eit

\section{Effective Scenarios with a Unified Cosmic Dark Sector} 
\label{sec:BEC-DS}

Since our interest in this paper is in the late-time cosmological evolution in 
the standard FRW framework, we resort to a scenario in which the universe is 
assumed to be constituted only by the BEC, which emulates a dust-like fluid of 
energy density given by Eq.\,(\ref{rhoB}), and the visible matter, also in the form of a dust (of baryons) of energy density 
\be \label{rhob}
\rbt(t) =\, \rbp\, a^{-3}(t) \,\,,
\ee 
where $\rbp$ denotes the corresponding value at the present epoch ($t = \tp$,
whence $a = 1$). 

\subsection{BEC cosmological equations}
\label{sec:BEC-Cosm-eqs}

Eqs.\,(\ref{rhoB}) and (\ref{rhob}) imply that the Raychaudhuri equation 
(\ref{QRE0}) can be written as
\be \label{QRE1}
\fr{\ddot a(t)}{a(t)} =\, -\, \fr{\k^2 \big[\rbp +\, \rBp\big]}{6 \, a^3(t)} 
+\, \fr 1 3 \, \UQ (t) \,\,.
\ee
While this is identified as the second Friedmann equation (\ref{cosm-eq2}) in the semi-classical picture, we have the total energy density of the universe 
expressed as
\be \label{rhotot}
\r\,(t) =\, \fr{\rbp +\, \rBp}{a^3(t)} -\, \fr 2 {\k^2} \, \UQ (t) -\, 
3 p\,(t) \,\,.
\ee
Note that $\UQ (t)$ is the background (or homogeneous) level 
quantum potential, mentioned above, i.e. the homogeneous part 
of the expression (\ref{QP}). 
For the BEC wave-function (\ref{WF1}), with the wave amplitude 
$R(t)$ given by Eq.\,(\ref{Rt}), this turns out to be
\be \label{QP1}
\UQ(t) =\, \fr{24 \hbar^2}{m^2 \s^4} \le[1 + \fr{\k^2 \s^2}{12} 
\le\{\Big[1 + \fr{3 \k^2 \s^2} 8 \, a(t) \, p'(t)\Big] \r(t) 
- 3 p(t)\ri\}\ri] \,,
\ee
where the prime $\{'\}$ denotes $d/da$ (see ref.
\cite{DSS-QB}
for the detailed steps in the computation).

In fact, rather conveniently, one may resort to the following 
form of the quantum potential, obtained as a function of the 
scale factor $a$, by substituting Eq.\,(\ref{rhotot}) in 
Eq.\,(\ref{QP1}) and re-arranging the terms:
\be \label{QP2}
\UQ(a) =\, \fr{6 \,\a^2}{1 +\, \a^2 \s^2 \,f(a)} \bigg[1 \,+\, \fr{\k^2 \s^2}{12} \bigg\{\le[\rbp +\, \rBp\ri] \fr{f(a)}{a^3} 
-\, 3 \le[1 + f(a)\ri] p(a)\bigg\}\bigg] \,, 
\ee
%
where 
\be \label{alpha}
\a =\, \fr{2 \hbar}{m \s^2} \,,
\ee
and 
\be \label{fa}
f(a) :=\, 1 +\, \fr{3 \k^2 \s^2 \, a \,p'(a)} 8 
\ee 
are, respectively, a constant of dimension $\k^{-1}$, and a dimensionless function of $a$.

On the other hand, substituting Eq.\,(\ref{rhotot}) in the conservation 
relation (\ref{consv-eq}) one derives the equation
\be \label{pU-eq}
\fr 2 a \, \Big[a^3 \UQ(a)\Big]' +\, 3 \k^2 \Big[a^2 p(a)\Big]' =\, 0 \,,
\ee
which requires to be solved simultaneously with Eq.\,(\ref{QP2}) in 
order to determine the Hubble rate $H(a)$, and hence the cosmic expansion 
history.
This is nevertheless a difficult proposition in general, since the potential 
$\UQ(a)$ in Eq.\,(\ref{QP2}) is given in terms of both $p(a)$ and $p'(a)$, 
as is evident from the definition (\ref{fa}) of the function $f(a)$. 
Substituting for $\UQ(a)$ in the first order differential equation 
(\ref{pU-eq}), one therefore finds a high degree of non-linearity of the 
resulting higher order differential equation for $p(a)$.

\subsection{Exact solution describing an effective $\L$CDM evolution}
\label{sec:BEC-LCDM}

Quite remarkably, however, an exact solution can be obtained simply by 
resorting to the ansatz 
\be \label{pr-const} 
p =\, -\, \L \,\,,
\ee 
where $\L$ is a (hitherto unspecified) constant
\cite{db2,db3,DSS-QB}. 
%
Such an ansatz is of course legitimate, and more importantly, the simplest possible one that renders the function $f$ in Eq.\,(\ref{fa}) just a constant ($= 1$), and thereby leads to a great deal of technical simplification while looking for a cosmological solution, in the sense that Eq.\,(\ref{QP2}) becomes an algebraic equation, instead of being a differential equation when $p' \neq 0$. Consequently, there is no non-linearity in Eq.\,(\ref{pU-eq}) when Eq.\,(\ref{QP2}) is substituted in it, and this really works\,! 
%
One can easily verify that the corresponding cosmological solution is the one that describes an effective $\L$CDM evolution, since the total energy density of the universe, given by Eq.\,(\ref{rhotot}), gets decomposed as
\be \label{LCDM-rho}
\rho(a) =\, \rmt(a) +\, \L \,, 
\ee 
where 
\be \label{LCDM-rm}
\rmt(a) =\, \rmp a^{-3} \,, 
\ee 
can be treated as the total dust-like matter density (with value $\rmp$ at the present epoch), and $\L$ as the effective DE density in the form of a cosmological constant. 

Now, it is natural to identify the total effective CDM density as
%
\be \label{LCDM-rcdm}
\rct(a) =\, \rmt(a) -\, \rbt(a) \,,
\ee 
where $\rbt = \rbp a^{-3}\,$ is the visible (baryonic) matter density. However, quite interestingly, neither $\rct$ is equal to the BEC density $\rBt = \rBp a^{-3}$, nor the cosmological constant $\L$ equals the quantum potential $\UQ$ (up to an appropriate dimensional scaling). 
To be more specific, the constancy of pressure (\ref{pr-const}), which implies $f = 1$, makes Eqs.\,(\ref{QP2}) and (\ref{pU-eq}) easily solvable, simultaneously, to yield $\UQ$ as the sum of a constant and a term proportional to $a^{-3}$. While the latter contributes to the total dust-like matter density of the universe, $\rmt$, the former (i.e. the constant part of $\UQ$) is equal to $\k^2 \L$, under the stipulation 
\be \label{Lambda}
\L =\, \fr{6 \, \k^{-2} \a^2}{1 -\, 2 \a^2 \s^2} \,\,. 
\ee
Explicit computations consequently reveal that $\rmp$ is not equal to $\rbp + \rBp$, i.e. the sum of the (visible) baryon and BEC densities at the present epoch. Instead, one can express it as
\cite{DSS-QB}
\be \label{LCDM-rm0}
\rmp \,= \le(1 -\, \vep\ri) \le[\rbp +\, \rBp\ri] \,, 
\ee
upon defining a dimensionless and positive definite constant parameter
\be \label{eps}
\vep \,:=\, \fr{\a^2 \s^2}{1 \,+\, \a^2 \s^2} 
\,=\, \fr{\Hp^2 \,\s^2 \,\OLp}{2 \,+\, 3 \Hp^2 \,\s^2 \,\OLp} \,\,,
\ee 
where $\OLp$ is the value of the $\L$-density parameter $\, \OLt = 
\L/\rho \,$ at the present epoch\footnote{Note that the last 
equality in Eq.\,(\ref{eps}) is obtained by eliminating $\a$ using 
Eq.\,(\ref{Lambda}), and from the definition $\OLp = \L/\r_{_0}$,
where $\r_{_0}^2 = 3 H_0^2/\k^2$ is the critical density at the 
present epoch.}.
 
The BEC density $\rBt$ is thus not accountable for the effective CDM 
density $\rct$ in entirety. In fact, from Eqs.\,(\ref{LCDM-rcdm}) and 
(\ref{LCDM-rm0}) it follows that $\rct$ is always less than $\rBt$, with 
the value at the present epoch determined as
\be \label{LCDM-rc0}
\rcp \,\equiv\, \rmp -\, \rbp = \le(1 -\, \vep\ri) \rBp -\, \vep \, 
\rbp \,\,.
\ee

\subsection{Quantum back-reaction and the BEC mass bound}
\label{sec:BEC-QB}

The net reduction in $\rcp$ from $\rBp$ is certainly a quantum effect, in 
the sense that it can be interpreted as the {\em quantum back-reaction} 
(QB) of the entire bulk of physical constituents of the universe (i.e. the 
BEC and the visible baryons) on the metric structure of space-time, by 
virtue of the quantum correction to the Raychaudhuri-Friedmann equation. 
The parameter $\vep$ (defined above) is therefore dubbed as the QB parameter 
in ref.
\cite{DSS-QB}. 
An estimation of this parameter, or at least the assertion of the feasible 
range of its values, is crucial for constraining the BEC mass parameter 
which can be expressed as
\be \label{BEC-mass}
m \,=\, \fr{m_{_H}}{2 \vep} \sq{\le(1 -\, \vep\ri) \le(1 -\, 3 \vep\ri) 
\OLp} \,\,,
\ee 
by recalling that $\,\a = 2 \hbar/(m \s^2)\,$ and using Eq.\,(\ref{eps}). 
Here, $\, m_{_H} = 2 \sq{2} \,\hbar \Hp \,$ is regarded as the `Hubble value' 
of the BEC mass, which one obtains in a Newtonian cosmological formulation, 
and under the consideration that all the bosons in the BEC are ultralight and 
neutral (for e.g. gravitons), with the Gaussian spread of the corresponding 
wave-function taken as $\s \simeq \Hp^{-1}$. As the Hubble constant $\Hp 
\simeq 10^{-42}\,$GeV, in the units $c = 1 = \hbar$, one gets $m_{_H} \simeq 
10^{-32}\,$eV 
\cite{db2,db3,DSS-QB}.

Now, the parameter $\vep$ being positive definite, its feasible range can be obtained from certain physical considerations: 
\bit 
\item Firstly, the positive definiteness of $\rmp$, and that of $\rbt$ and 
$\rBt$, implies $\vep \in (0,1)$, as is evident from Eq.\,(\ref{LCDM-rm0}).
\item Secondly, the real-valuedness of $m$, given by Eq.\,(\ref{BEC-mass}), 
restricts the domain further to $\vep \in \le(0, \rfra 1 3\ri)$.
\item Finally, an even tighter upper bound on $\vep$ can be obtained from 
the legitimate demand that the BEC density $\rBt$ must not exceed the total 
matter density $\rmt$ at any phase of evolution of the universe. This 
amounts to imposing the condition $\, \rBp < \rmp \,$ at the present epoch 
$t = \tp$. Hence by Eq.\,(\ref{LCDM-rm0}), 
\be \label{eps-bound}
\vep \,<\, \fr{\Obp}{1 -\, \OLp +\, \Obp} \,\,,
\ee
where $\Obp$ is the value of the baryon density parameter $\Obt = \rbt/\rho\,$ 
at $t = \tp$. Using therefore the best-fit $\OLp$ and $\Obp$ estimations from, 
say, the combined analysis of Planck 2018 (TT,TE,EE+lowE), Lensing and Baryon 
Acoustic Oscillations (BAO) data
\cite{Planck18-CP},
for the base $\L$CDM model, one finds $\vep \leq 0.136$
\cite{DSS-QB}. 
\eit 
Note that we have excluded the possibility $\vep = 0$ right from the onset, 
as otherwise, by Eq.\,(\ref{BEC-mass}), the mass parameter $m$ would blow up. 
On the other hand, the small upper limit of $\vep$ ($= 0.136$, found above) 
implies that $m$ is higher than the Hubble value $m_{_H} \simeq 10^{-32}$~eV 
by at least a numerical factor of $2.18$ or so. However, this bound is too 
loose to be considered as a useful constraint when it comes to realize whether 
the BEC behaves primarily as a self-consistent (and dominating) DM component 
of the universe, or it is so ultralight to be comparable only to a scalar 
field DE candidate. 

Let us point out here that a dominant DM species is desired to have a mass 
higher than $m_{_H} \simeq 10^{-32}$~eV by a few orders of magnitude, in 
order to be free from the inconsistencies with observations at smaller 
(galactic) scales of the order $\lesssim 10$~Kpc (see for e.g.
\cite{ABP-CDMprob,PFMK-CDMprob,PD-CDMprob,BB-CDMprob}
and the references therein). By this, we refer in particular to the mass 
bounds for which the small scale structure (DM halo or sub-halo) formation 
gets suppressed, or at least not invoke any significant effect at larger 
scales. 
One may reckon the {\it fuzzy} DM hypothesis pertaining to ultralight 
bosonic particles of masses $\lesssim 10^{-22}$ eV, but with large 
de Broglie wavelengths that give the effective Jeans scale, below which 
the stability of the DM halo is ensured according to the uncertainty 
principle 
\cite{HBG-FDM}.
However, the viability of the models with ultralight bosonic dark matter 
(UBDM) candidates, such as {\it axions}, depend on severe constraints on 
the corresponding mass value from, e.g., the cosmic microwave background 
(CMB), pulsar timing array (PTA), black hole super-radiance, as well as 
the Lyman-$\a$ forest power flux observations of distant quasars (see 
\cite{MH-UBDM,fer-UBDM,marsh-UBDM}
for recent reviews). In fact, the viable mass of a dominating axionic ULDM 
species is inferred to be no less than $m \simeq 10^{-24}\,$eV, i.e. about 
$1/100$-th of the fuzzy DM benchmark ($10^{-22}\,$eV) quoted above
\cite{MN-UBDM,APYMB-UBDM,CAMD-UBDM,HOTW-UBDM,HGMF-UBDM}.
On the other hand, the mass of an axion-like DE candidate is argued to be 
within the proximity of $m_{_H} \simeq 10^{-32}$~eV, or is deviated by one 
to two orders of magnitude at most
\cite{MH-UBDM,fer-UBDM,marsh-UBDM}.

Nevertheless, these mass bounds, which are generally obtained under 
the presumption of independently evolving DM and DE, may not be 
amenable in an interacting or unified scenario, such as the one 
emerging from the BEC cosmological formulation in this paper. One
may still argue that the small scale crisis for the 
CDM can be ameliorated by the BEC if its mass $m$ is as high as 
$10^{-24}\,$eV or so. By Eq.\,(\ref{BEC-mass}), this corresponds to 
$\vep \simeq 10^{-8}$, which is of course well within the bound 
$\vep \lesssim 0.136$ obtained for the $\L$CDM solution of the BEC 
cosmological equations. However, this bound is too loose to allow
even a value of $m$ within the close proximity of $m_{_H} \simeq 
10^{-32}$~eV, which may thereby imply that the BEC acts predominantly
as a DE, rather than a DM candidate. Also, there is no concreteness 
in the above argument, since after all, a negligible quantum 
back-reaction ($\vep \simeq 10^{-8}$) does not imply the interaction 
of the effective CDM and DE components to be so. It is therefore 
imperative to estimate directly (and robustly) the parameter $\vep$, 
and hence the mass ratio $m/m_{_H}$, using the observational data, 
rather than looking for a crude bound merely from a physical 
consideration.
However, such an estimation, in an analysis with the commonly used 
SN-Ia and Hubble data, is untenable for the $\L$CDM solution, as 
the corresponding expression for reduced Hubble parameter $H(a)/\Hp$  
does not involve the parameter $\vep$ explicitly. Hence there is a 
need to look for a more general solution of the BEC cosmological 
equations, which may yield $\L$CDM in a limit. Despite the high 
degree of analytical difficulty,  we work out 
a way of obtaining such a solution, by making the key utilization of 
the smallness of $\vep$, as described in the next subsection.


\subsection{A unified scenario of dark matter and a dynamical dark energy} 
\label{sec:BEC-dynDE}

Let us recall the expression (\ref{QP2}) for the quantum potential 
$\UQ \,$, and re-write it as follows, in terms of the 
parameters $\L$ and $\vep$ defined by Eqs.\,(\ref{Lambda}) and 
(\ref{eps}) respectively:
\be \label{QP4}
\UQ(a) =\, \fr{\le(1 - 3\vep\ri) \k^2 }{1 - \vep \le[1 - f(a)\ri]} \bigg[\L +\, \fr \vep {2 \le(1 - 3\vep\ri)} \bigg\{\le[\rbp + \rBp\ri] \fr{f(a)}{a^3} 
-\, 3 \big[1 + f(a)\big] p(a)\bigg\}\bigg] \,,
\ee
where 
\be \label{f}
f (a) =\, 1 +\, \fr{9 \vep \, a \, p'(a)}{4 \le(1 -\, 3\vep\ri) \L} 
\ee
is the function given by Eq.\,(\ref{fa}), re-expressed accordingly.

Plugging Eq.\,(\ref{QP4}) in the differential equation (\ref{pU-eq}), 
we can in principle solve for the pressure $p\,(a)$, and thereby 
determine the expansion history of the universe ($\r\,(a)$ or $H(a)$) 
using Eq.\,(\ref{rhotot}). However, given the complicated form of 
Eq.\,(\ref{QP4}), it is very difficult to get the general solution 
analytically. Numerical techniques, on the other hand, require 
certain fiducial parametric range specifications, i.e. suitable priors 
on some of the parameters in the set $\le(\rbp, \rBp, \L, \vep\ri)$, 
apart from the appropriate usage of the initial conditions. Such 
priors are bound to have some arbitrariness in their selection, and 
the numerical solution would be far from being a general solution 
under the given circumstances. Nevertheless, deciding on the extent 
of the generality is our prerogative, in the sense that we forbid any 
large deviation from $\L$CDM, in view of the general observational 
concordance on the latter. To be more specific, the general solution, 
or any particular solution more general than $\L$CDM, essentially 
implies a dynamical evolution of the effective DE component. However, 
we demand the dynamics to be {\em mild} enough, so that the cosmic 
parametric estimates do not breach the corresponding $\L$CDM margins 
very significantly. Now, with such a demand, we may still reckon the 
effectiveness of the quantum back-reaction we have had for the 
$\L$CDM case in the previous subsection, and the smallness of the 
respective parameter $\vep$ therein. It then suffices to look for a 
solution of the cosmological equations, which may be regarded as the 
general solution, but only up to, say, the leading order in the 
powers of $\vep$. This is the very reason why we have recast $\UQ$ 
explicitly in terms of $\vep$ and $\L$ in Eq.\,(\ref{QP4}). The 
following is the power series expansion for the same:
\be \label{QP-series}
\UQ(a) =\, \k^2 \le[\L +\, \vep \le\{\fr{\rbp +\, \rBp}{2 a^3}
\,-\, 3 \Big[p (a) +\, \L\Big]\ri\} +\, \cO \big(\vep^2\big)\ri] \,.
\ee
This is certainly a lot simpler to handle, compared to the full 
expression (\ref{QP4}). 

Keeping terms up to $\cO \big(\vep\big)$ in $\UQ$, we solve the 
differential equation (\ref{pU-eq}) to obtain
\be \label{p-approx}
p (a) \,\approx\, \pp a^{-2s} -\, \L \Big(1 -\, a^{-2s}\Big) \,\,,
\ee 
where $\pp = p\big\vert_{t = \tp}$ is the pressure at the present 
epoch, and 
\be \label{s}
s =\, \fr{1 - 3\vep}{1 - 2\vep} \,\approx\, 1 - \vep \,.
\ee 
Substituting Eq.\,(\ref{p-approx}) back in Eq.\,(\ref{QP-series}),
we get the approximated quantum potential
\be \label{QP-approx}
\UQ(a) \,\approx\, \k^2 \le[\L +\, \vep \le\{\fr{\rbp +\, \rBp}{2 a^3} 
\,-\, \fr{3 \le(\pp +\, \L\ri)}{a^{2s}}\ri\}\ri] \,,
\ee
which when plugged, alongwith Eq.\,(\ref{p-approx}), in Eq.\,(\ref{rhotot}), leads to the approximated total energy density
%
\be \label{rho-approx}
\r\,(a) \,\approx\, \fr \rmp {a^3} \,+\, \L \,-\, \fr{3 \le(1 - 2\vep\ri) \le(\pp +\, \L\ri)}{a^{2s}} \,, 
\ee 
%
where the first term is the total dust-like matter density of the universe, whose value at the present epoch is $\rmp = \le(1 - \vep\ri) \le(\rbp + \rBp\ri)\,$ [{\it cf}. Eq.\,(\ref{LCDM-rm0})]. The other terms in Eq.\,(\ref{rho-approx}) can collectively be treated as the total effective DE density $\rx(a)$, 
%
%
%
whence the corresponding (DE) equation of state (EoS) parameter can be expressed as
\be \label{DE-wx}
\wx (a) \,:=\, \fr{p\,(a)}{\rx (a)} \,\approx\, -\, 1 \,+\, \fr{2 \le(1 - 3\vep\ri) \le(\pp +\, \L\ri)}{3 \le(1 - 2\vep\ri) \le(\pp +\, \L\ri) -\, \L a^{2s}} \,.
\ee 
The overall cosmological evolution is therefore quintessance-like ($\wx > -1$), as long as $\pp > -\L$.

Substituting Eq.\,(\ref{rho-approx}) in the Friedmann equation (\ref{cosm-eq1}) we have
\be \label{ratHub}
\le[\fr{H(a)}{\Hp}\ri]^2 \,\approx\, \fr{\Omp}{a^3} +\, \OLp -\, \fr{3 \le(1 - 3\vep\ri) \le[\OLp +\, \sw_{_0}\ri]}{a^{2s}} \,,
\ee
where $\sw_{_0}$ is the value of the total EoS parameter of the system, $\sw = p/\r$, at the present epoch ($a = 1$), $\Omp$ and $\OLp$ are, respectively, the matter density parameter and the $\L$-density parameter at $a = 1$, and $\Hp$ is the Hubble constant. In fact, since $H = \Hp$ at $a = 1$, we have the constraint
\be \label{w0}
\sw_{_0} \approx\, -\, \fr{1 -\, \Omp +\, 2 \le(1 - 2\vep\ri) \OLp}{3 \le(1 - 2\vep\ri) } \,.
\ee
Using this, Eq.\,(\ref{ratHub}) can be re-expressed as
\be \label{ratHub-1}
\le[\fr{H(a)}{\Hp}\ri]^2 \,\approx\, \fr \Omp {a^3} +\, \OLp + \fr{1 - \Omp - \OLp}{a^{2s}} \,,
\ee
or more conveniently, as 
\be \label{ratHub-2}
\le[\fr{H(z)}{\Hp}\ri]^2 \,\approx\,
\Omp \le(1 + z\ri)^3 +\, \OLp + \le[1 -\, \Omp -\, \OLp\ri] 
\le(1 + z\ri)^{2 \le(1 - \vep\ri)} \,,
\ee
where $z = \le(a^{-1} - 1\ri)$ denotes the redshift, and we have used the relation $s \approx 1 - \vep\,$ [{\it cf}. Eq.\,(\ref{s})]. 

The crucial point to note here is that the parameter $\OLp$, which arises due to the quantum correction to the Raychaudhuri equation, is a free parameter, and not to be confused with the DE density parameter at the present epoch, $\, \OXp := \le[\rx/\rho\ri]_{z=0}\,$, which is by definition equal to $\, 1 - \Omp$.

It is also worth pointing out that a full perturbative approach, i.e. the consideration of the series expansion not only of $\UQ(a)$, but also of $p(a)$ and $\rho(a)$, in powers of $\vep$, leads to 
\bea
\UQ(a) &\approx & \k^2 \le[\L +\, \vep \le\{\fr{\rbp +\, \rBp}{2 a^3} 
\,-\, \fr{3 \le(\pp +\, \L\ri)}{a^2}\ri\}\ri] \,, \label{QP-fapprox} \\
p (a) &\approx & -\, \L +\, \fr{\le(\pp + \L\ri)\le(1 + 2 \vep N\ri)}
{a^2} \,, \label{p-fapprox}\\
\r\,(a) &\approx & \fr \rmp {a^3} +\, \L -\, \fr{3 \le(\pp + \L\ri)
\le[1 +\, 2 \vep \le(N - 1\ri)\ri]}{a^2} \,, \label{rho-fapprox}
\eea
where $N = \ln a$. However, up to $\cO \big(\vep\big)$, the solution set 
(\ref{p-approx})\,--\,(\ref{rho-approx}) is more general, and therefore we shall stick to it in what follows.


\section{Estimation of Cosmological Parameters and BEC Mass} 
\label{sec:BEC-est}

Let us proceed to carry out the statistical estimation of the parameters $\Omp$, $\OLp$ and $\vep$ in Eq.\,(\ref{ratHub-2}), and also the reduced Hubble constant $h= \Hp/\!\le[100 \, \mbox{Km s}^{-1}\, \mbox{Mpc}^{-1}\ri]$ (where relevant), using the well-known Metropolis-Hastings algorithm for the Markov-chain Monte Carlo (MCMC) random probabilistic exploration. For this purpose, we consider the following (reasonably wide) pre-assigned domain, or prior range, of those parameters:
\bea \label{priors}
&&\Omp \in [0.05,0.5] \,, \qquad \OLp \in [0.45,0.95] \,, \nn\\ 
&&\vep \in [0,0.25] \,, \qquad h \in [0.55,0.85] \,,
\eea
and use the SN-Ia Pantheon (binned) data-set
\cite{scol,panth}
and its combination with observational Hubble $(H(z))$ data-set 
\cite{MPCJM-OHD,YRW-OHD,RDR-OHD}.
While the former refers to a list of $40$ data-points showing the 
redshift $z$ versus the distance modulus $\m(z)$ for well-observed 
supernovae Ia, the latter contains $30$ data-points listing $z$ 
versus $H(z)$ direct measurements using cosmic chronometry detailed 
in ref.
\cite{mors}.

The standard technique involves the minimization of 
\be \label{chi2-def}
\chi^2 :=\, \sum_{i,j} \D V_i \cdot C^{-1}_{ij} \cdot \D V_j \,\,,
\ee
where $\D V_i = V_{\mbox{\scriptsize obs}} (z_i) - V (z_i) \,$ denotes the difference between the observed and the theoretically predicted values of the observable $V$ under consideration (i.e. $\m$ or $H$) at a given redshift $z_i$, and $C_{ij}$ is the covariance matrix for the data-points.
%
\begin{table}[ht]
\caption{\footnotesize{Best fit values and(or) $1\s$ 
confidence limits of parameters $\Omp$, $\OLp$, $h$ and 
$\vep$, along with the minimized $\chi^2$, for estimations 
using the Pantheon and Pantheon$+H(z)$ datasets.}}
{
\begin{center}
\renewcommand{\arraystretch}{1.5}
{
\begin{tabular}{||c||c|c|c|c||c||}
\hline
 & \multicolumn{4}{c||}{Parametric estimations} & \\
Observational  &  \multicolumn{4}{c||}{\footnotesize (Best fit \& 
$1\s$ limits)} & ~ $\chi^2_{_{\mbox{\scriptsize min}}}$ \\
\cline{2-5}
%
datasets & $\Omp$ & $\OLp$ & $h$ & $\vep$ &  \\
\hline\hline
{\small Pantheon} & $0.2886^{+0.0353}_{-0.0426}$ & 
$0.7018^{+0.0656}_{-0.0531}$ & -- & $< 0.0693$ &  $48.4279$  \\
\hline
{\small Pantheon$+H(z)$ } & $0.2873^{+0.0312}_{-0.0364}$ & 
$0.6916^{+0.0478}_{-0.0446}$  & $0.6839^{+0.0239}_{-0.0273}$ & 
$< 0.0675$ & $63.4063$ \\
\hline\hline
\end{tabular}
}
\end{center}
\label{BEC-EstTable} }
\end{table} 
%
In principle, one may carry out the minimization of $\chi^2$ with respect to each of the parameters involved. However, for perspicuity in the approach, while using the Pantheon data-set, we prefer to follow the two-step procedure that is usually adopted --- integrating out or marginalizing first the two superfluous parameters that call into play, viz. the absolute magnitude $M$ and the Hubble constant $\Hp$ (or $h$), and then minimizing the resulting $\chi^2$ with respect to the three parameters of significance, viz. $\Omp$, $\OLp$ and $\vep$. While using the Pantheon$+ H(z)$ data-set though, $h$ has to be considered as a parameter of significance as well, and no longer a superfluous (or nuisance) parameter.

\begin{figure}[!htp]
\begin{subfigure}{\linewidth} \centering
   \includegraphics[height=3.75in,width=4.25in]{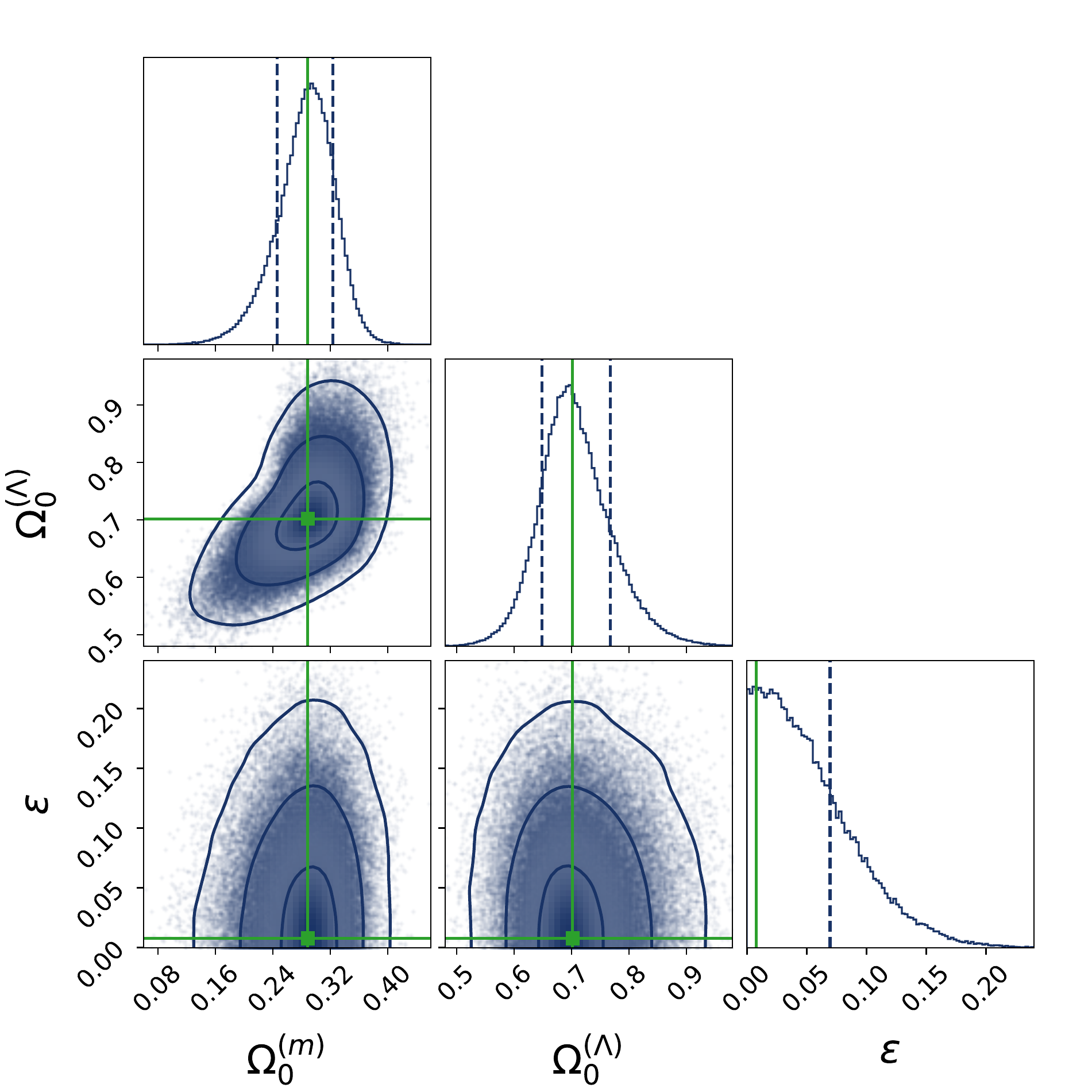}
   \caption{\footnotesize Two-dimensional posterior distribution using the 
   Pantheon data-set.}
   \label{BEC_Panth}
\end{subfigure} 
%
\begin{subfigure}{\linewidth} \centering
    \includegraphics[height=4.5in,width=5in]{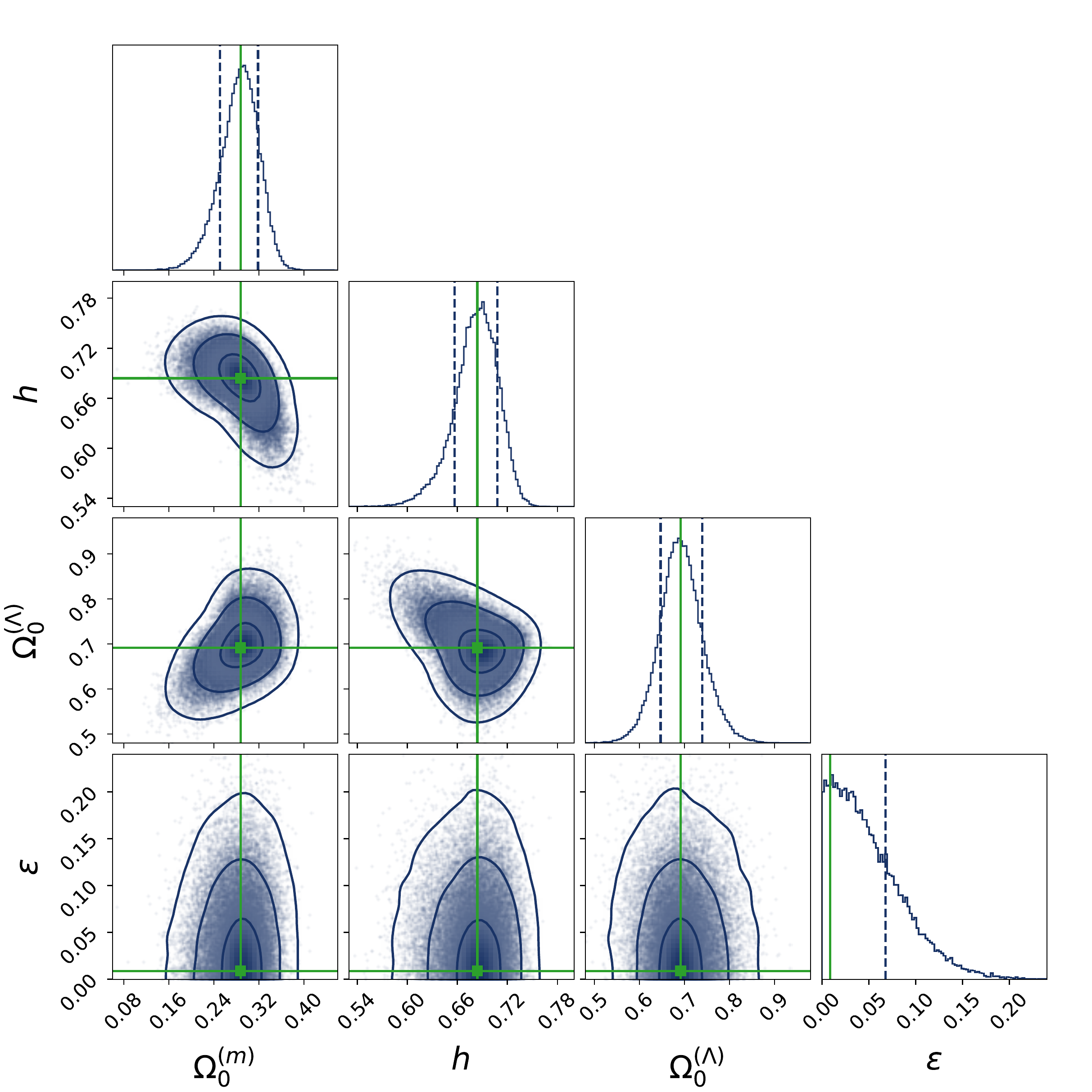}
    \caption{\footnotesize Two-dimensional posterior distribution using the 
    Pantheon$+H(z)$ data-set.}
    \label{BEC_PanHubb}
\end{subfigure} 
\caption{\footnotesize $1\s$-$3\s$ parametric contour levels for estimations using (a) Pantheon data and (b) Pantheon$+ H(z)$ data. The plot densities of the regions bounded by the 
contours are shown with the data point scatter, whereas the central 
solid dots indicate the median best-fits of the corresponding parameters.}
\label{BEC_fig1}
\end{figure}

Table\,\ref{BEC-EstTable} quotes the best fit values and $1\s$ 
confidence limits of such parameters, estimated using the Pantheon 
and the Pantheon$+ H(z)$ data, as well as the corresponding minimized 
$\chi^2$ values (denoted by $\chi^2_{_{\mbox{\scriptsize min}}}$).
Figs.\,\ref{BEC_fig1}\,(a) and (b), on the other hand, show the respective 
two-dimensional posterior distribution or the error contour levels, up to 
$3\s$. 
%
%
%

Quite understandably, with a larger number of data-points, the 
Pantheon$+ H(z)$ data-set constrains the parameters $\Omp$, $\OLp$ and 
$\vep$ more tightly than the Pantheon data-set. It is also worth noting 
that the best fit value of $\vep$ turns out to be quite insignificant 
(about three orders of magnitude below unity), which is why we have 
quoted only its $1\s$ margin in the Table\,\ref{BEC-EstTable}. 

\begin{figure}[!h]
\centering
\begin{subfigure}{0.48\linewidth} \centering
   \includegraphics[scale=0.5]{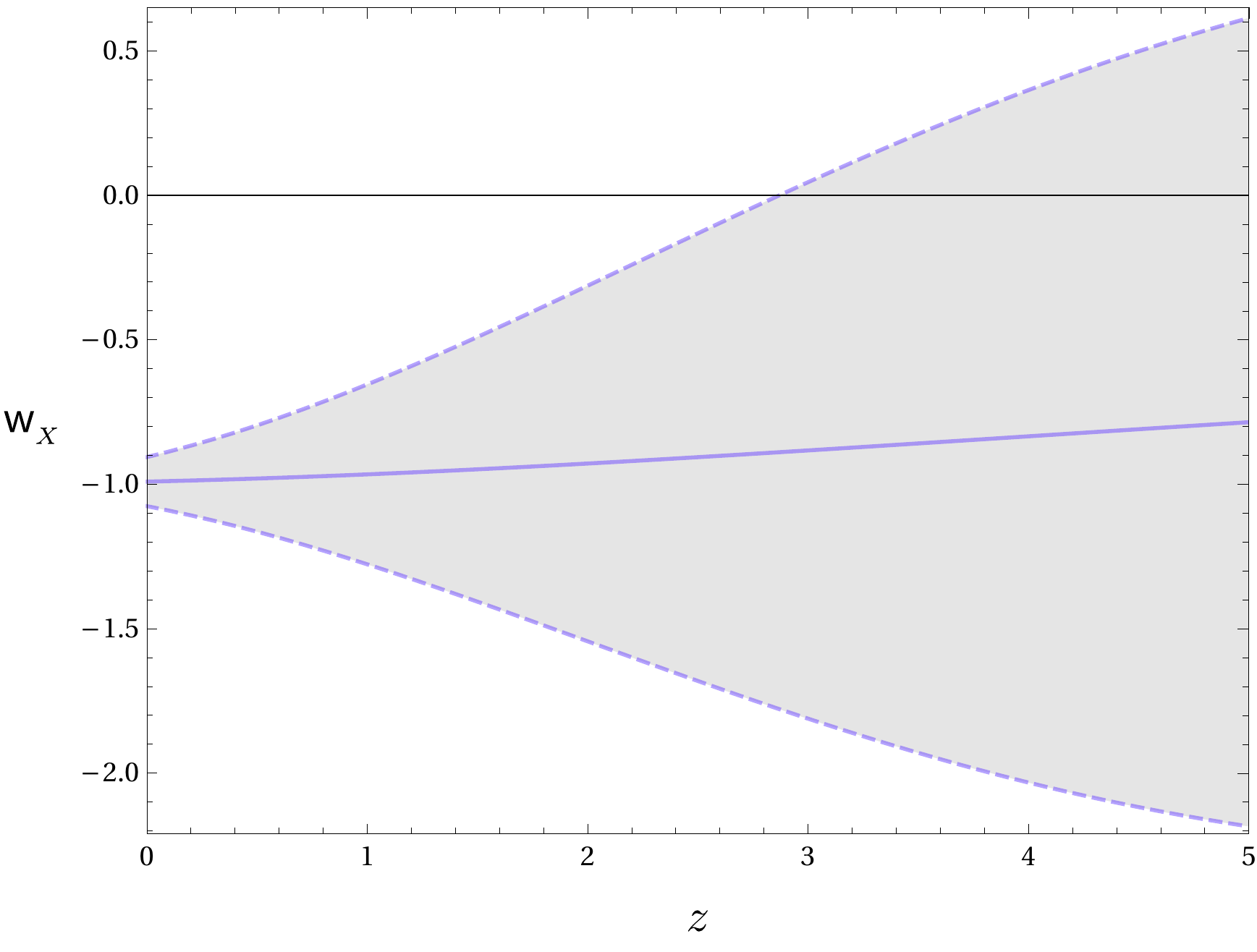}
   \caption{\footnotesize $\wx (z)$ from the Pantheon 
   data.} 
   \label{BEC_Panth_wx}
\end{subfigure} ~~
\begin{subfigure}{0.48\linewidth} \centering
    \includegraphics[scale=0.5]{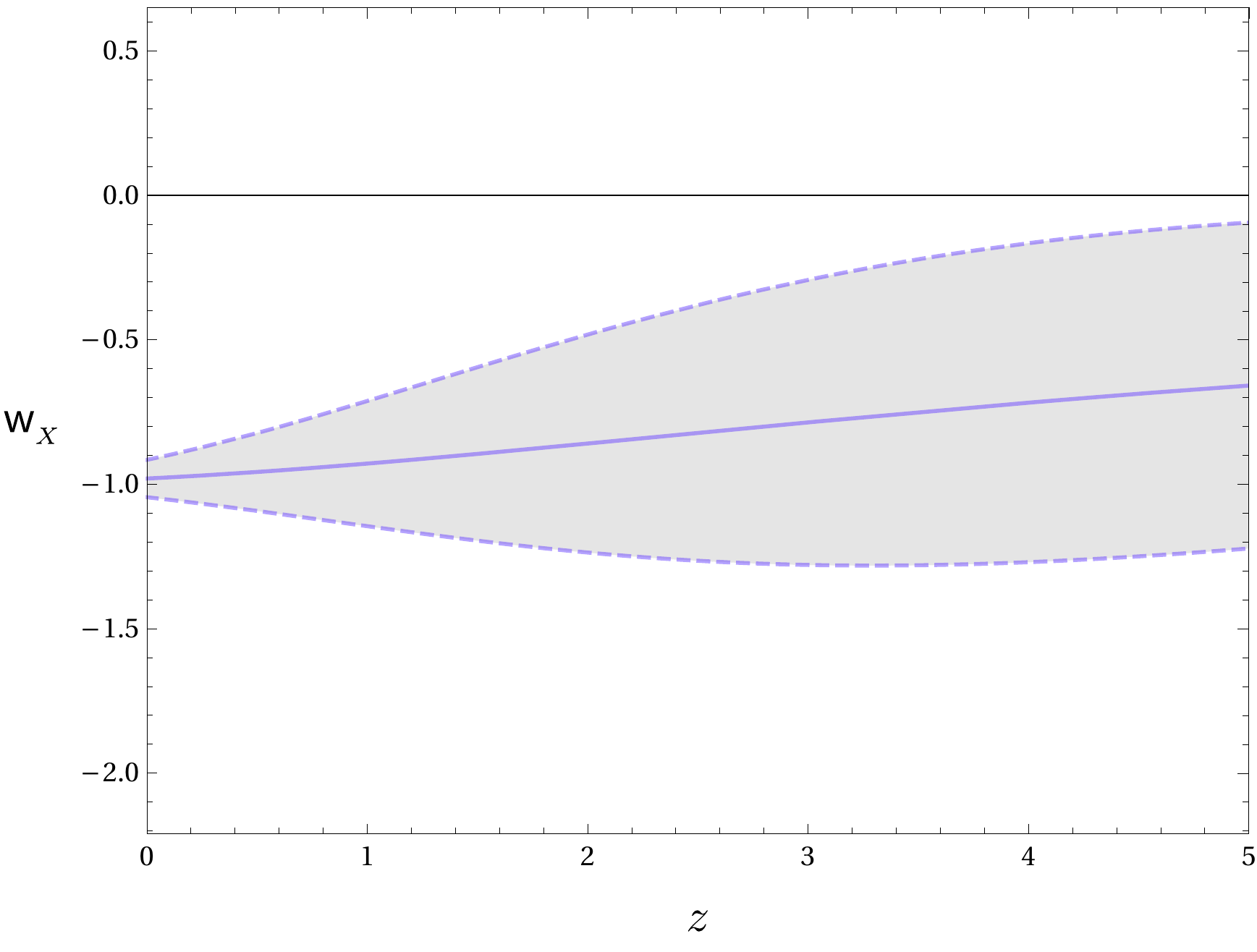}
    \caption{\footnotesize $\wx (z)$ from the 
    Pantheon$+ H(z)$ data.} 
    \label{BEC_PanHubb_wx}
\end{subfigure} 
\caption{\footnotesize Effective dark energy EoS parameter $\wx (z)$ 
(solid line) and its $1\s$ error limits (broken lines bounding the shaded regions) for estimations using (a) Pantheon data and (b) Pantheon$+ H(z)$ data.}
\label{BEC_fig2}
\end{figure}

On the whole, however, there is not much deviation from the effective 
$\L$CDM scenario we have had in subsection \ref{sec:BEC-LCDM}, as the 
corresponding parametric values turn out to be well within the respective 
$1\s$ domains obtained here, for the more general cosmological evolution 
described by Eq.\,(\ref{ratHub-2}). One may see this from the $1\s$ 
domains of the effective DE equation of state parameter $\wx$, derived 
using the parametric estimates in the Table\,\ref{BEC-EstTable} and 
shown in Figs.\,\ref{BEC_fig2}\,(a) and (b), for the two data-sets. 
%
%
\begin{figure}[!h]
\begin{subfigure}{0.495\textwidth} \centering
   \includegraphics[scale=0.5]{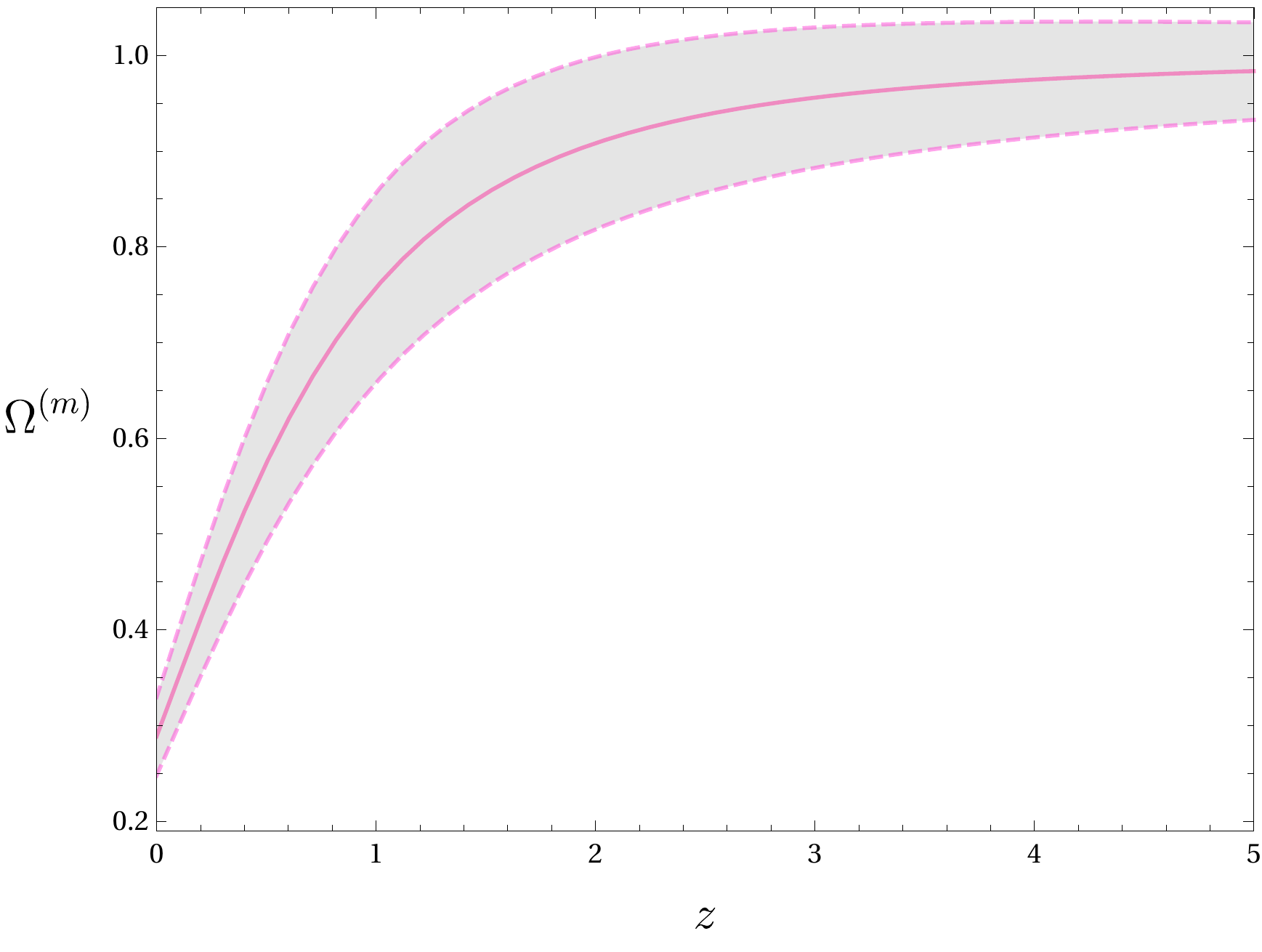}
   \caption{\footnotesize $\Omt (z)$ from the Pantheon 
   data.} 
   \label{BEC_Panth_Om}
\end{subfigure} ~
\begin{subfigure}{0.495\textwidth} \centering
    \includegraphics[scale=0.5]{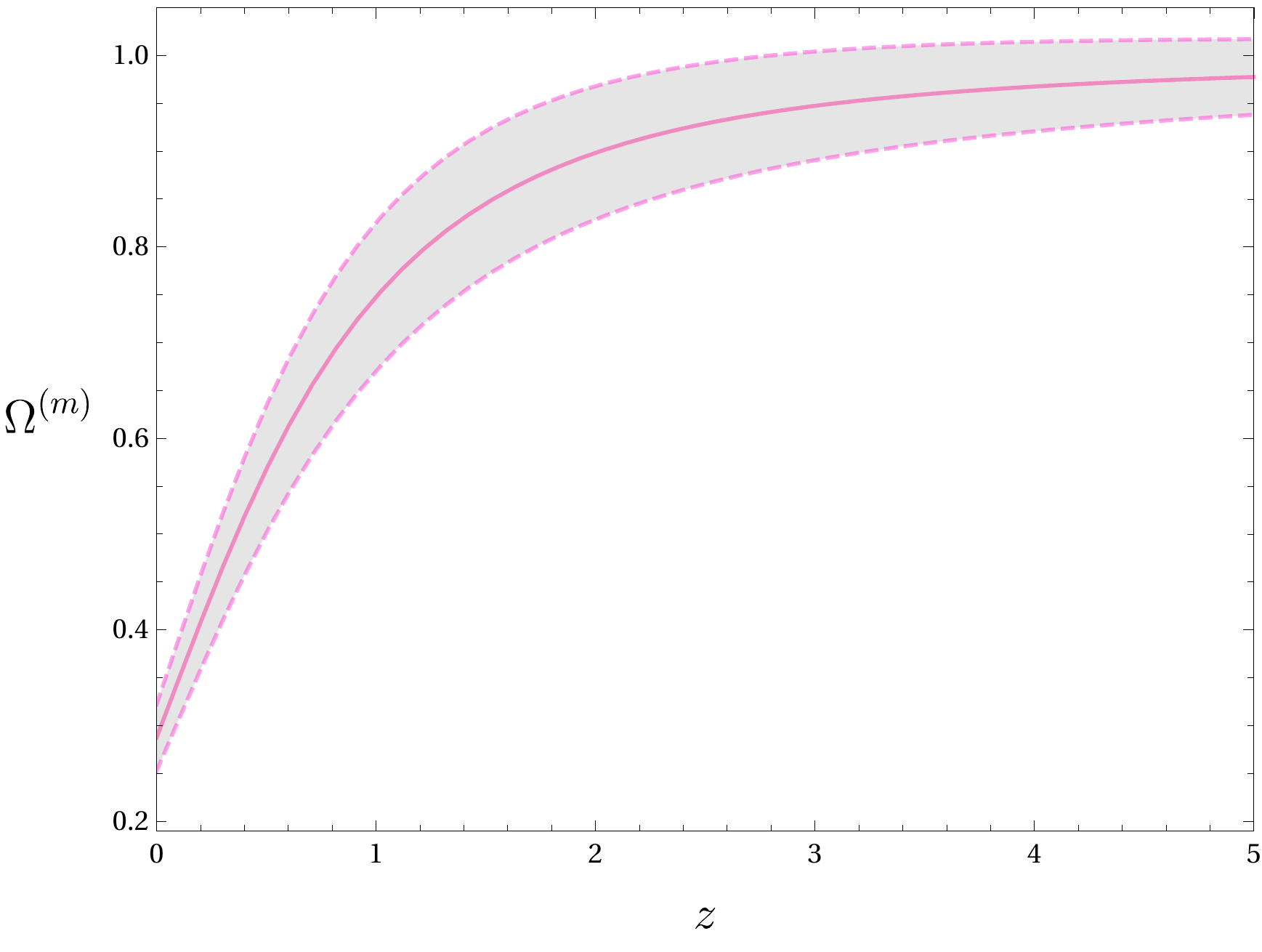}
    \caption{\footnotesize $\Omt (z)$ from the 
    Pantheon$+ H(z)$ data.} 
    \label{BEC_PanHubb_Om}
\end{subfigure} 
\caption{\footnotesize Effective total matter density parameter 
$\Omt (z)$ (solid line) and its $1\s$ error limits (broken lines bounding the shaded regions) for estimations using (a) Pantheon data and (b) Pantheon$+ H(z)$ data.}
\label{BEC_fig3}
\end{figure}
%
The $\L$CDM value $\wx = -1$ is well within such domains. However, the 
derived $1\s$ domains of the total matter density parameter $\Omt$ 
stretch beyond unity at higher redshifts ($z \gtrsim 2$), as shown in 
Figs.\,\ref{BEC_fig3}\,(a) and (b), for the respective data-sets. 
This is not unexpected though, from the point of view of our perception 
of the total matter density being that of the net effective dust-like 
matter content of the universe. In other words, whilst the visible 
matter being considered to be entirely pressure-less here, not all of 
the CDM (which is also pressure-less) is due to the BEC. The quantum 
back-reaction that calls into play, can effectively let $\Omt$ exceed 
unity at high $z$. Moreover, the very extrapolation of the $\Omt$ plots 
to high $z$ ($\gtrsim 2$) may not be quite reliable, given that the 
SN-Ia data are available only up to $z \lesssim 2$. In any case, 
$\Omt > 1$ at the $1\s$ level is nothing new in a unified picture,
because what are being interpreted here as the effective CDM and DE 
components, are nonetheless interacting. 

\begin{figure}[!h]
\centering
\begin{subfigure}{0.48\linewidth} \centering
   \includegraphics[scale=0.5]{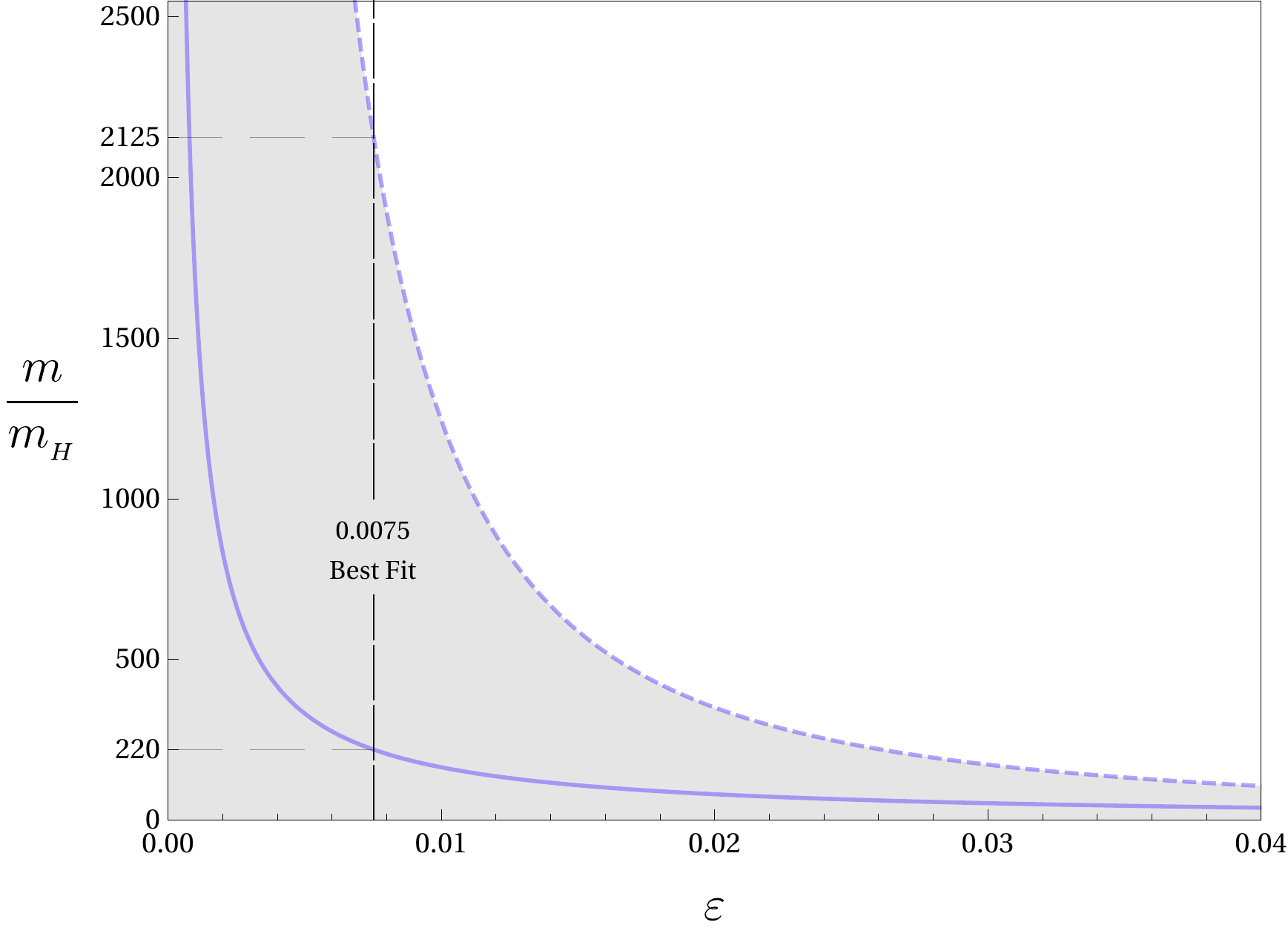}
   \caption{\footnotesize $m/m_{_H}$ (up to $1\s$) vs 
   $\vep$ for Pantheon.} 
   \label{BEC_Panth_M_e}
\end{subfigure} ~
\begin{subfigure}{0.48\linewidth} \centering
    \includegraphics[scale=0.5]{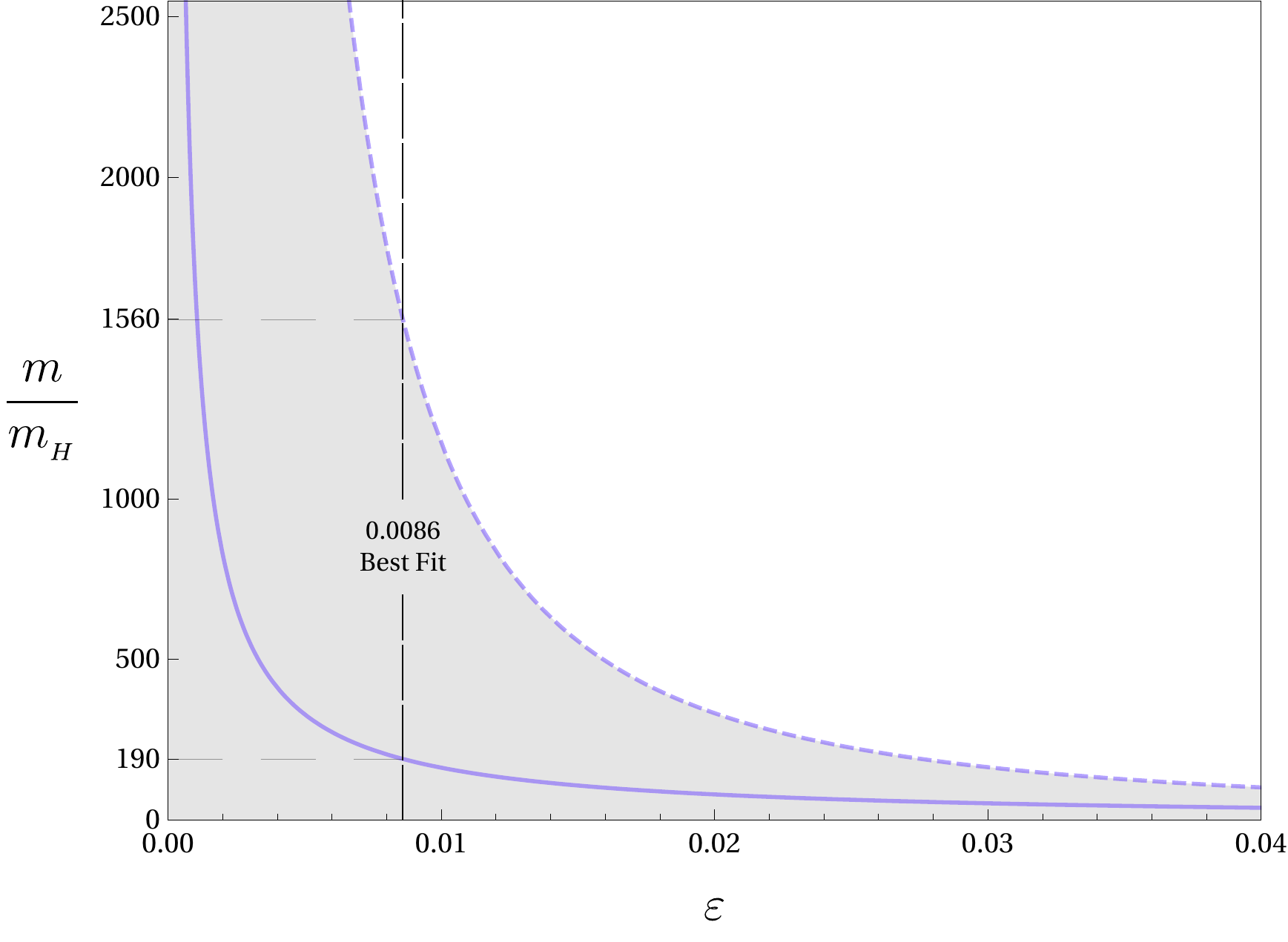}
    \caption{\footnotesize $m/m_{_H}$ (up to $1\s$) vs 
    $\vep$ for Pantheon$+ H(z)$.} 
    \label{BEC_PanHubb_M_e}
\end{subfigure} 
\caption{\footnotesize Variation of the rationalized BEC mass $m/m_{_H}$ with $\vep$, for estimation (up to $1\s$) using (a) Pantheon data and (b) Pantheon$+H(z)$ data. The solid lines and broken vertical lines denote the best 
fits of $m/m_{_H}$ and $\vep$ in the respective cases.}
\label{BEC_fig4}
\end{figure}

Figs.\,\ref{BEC_fig4}\,(a) and (b) show the $\vep$-variation of the best fit and $1\s$ error margins of the rationalized BEC mass $m/m_{_H}$, estimated using Eq.\,(\ref{BEC-mass}) and the cosmological parametric values quoted in Table\,\ref{BEC-EstTable}, for the two data-sets.
%
%
With decreasing $\vep$, the mass gets enhanced steeply, 
since by Eq.\,(\ref{BEC-mass}), $m/m_{_H}$ varies almost inversely as $\vep$, for small values of the latter.
Up to $1\s$, the mass enhancement is about three orders of
magnitude over $m_{_H} \simeq 10^{-32}$\,eV, which is still 
fairly in the range expected for a scalar field DE candidate 
(rather than that for an axion-like DM). 

\section{A BEC Dark Universe with a Spatial Curvature} 
\label{sec:BEC-kDS}

While the parametric estimations in previous section, for the cosmological solution obtained in subsection \ref{sec:BEC-dynDE}, show a very small upper limit of the BEC mass $m$, the question that naturally arises is {\em whether it would still be so, if the assumption of spatial flatness is relaxed}. After all, remember that it is the quantum back-reaction parameter $\vep$ which is solely responsible for the dynamics of the effective DE component of the universe. However, such a DE dynamics only mildly distorts the $\L$CDM evolution, because of the smallness of the $\vep$ that has been demonstrated in subsection \ref{sec:BEC-QB} from purely physical considerations. Therefore, since the smallness of $\vep$ holds the key here, it is natural to look for its overall effect without assuming the universe to be spatially flat beforehand. In fact, note that with the spatial flatness assumption, the estimated $\vep$, although small (up to $1\s$), is not negligible enough to enhance the BEC mass $m$ (which roughly varies as $\vep^{-1}$) from the Hubble value $m_{_H}$, to a great extent. As such, $m$ is left in the range of that for a scalar field DE, which is of course quite contrary to the common consideration of the BEC as a typical DM candidate. It is therefore imperative to examine whether this would still be the case, when the density parameter $\Okt = - k/(a^2 H^2)$, corresponding to the spatial curvature constant $k$, is not assumed to be ignorable at any epoch.

On the other hand, given the smallness of $\vep$, the approximate solution (\ref{rho-approx}) for the total energy density $\r$, in subsection \ref{sec:BEC-dynDE}, has the striking feature of distorting $\L$CDM only to the extent of that due to almost an effective spatial curvature (as noticed from the last term therein, which very nearly varies as $a^{-2}$, since the parameter $s \approx 1 - \vep\,$ [{\it cf}. Eq.\,(\ref{s})]). In other words, we almost have an effective $k\L$CDM evolution in the BEC cosmological setup, and that too when $k=0\,$! This is in fact even more clear from the full perturbative solution (\ref{rho-approx}) for $\r$, shown at the end of subsection \ref{sec:BEC-dynDE}. One is therefore urged further to examine the nature of the solutions of the Friedmann equations in such a setup, for a non-vanishing $k$, and see how the latter affects the already very nearly $k\L$CDM evolution found in its absence. 

A generalization of the BEC cosmological formalism in sections \ref{sec:BEC-Cosm} and \ref{sec:BEC-DS}, in presence of a non-zero $k$ or $\Okt$, is not an easy task though. In fact, given a specific functional form of the BEC wave-function, such as that in Eq.\,(\ref{WF1}), the computation of the quantum potential $\UQ$ is complicated by itself, as the general form of $\UQ$, given by Eq.\,(\ref{QP}), involves as many as four space-time derivatives --- two in the $\square$ operator and two outside of it. Nevertheless, for $k = 0$, one can have a good deal of simplification in such a computation, as the form of the (spatially flat) FRW line element is simple enough, especially in the Cartesian basis. For $k \neq 0$, the steps in the computation, although complicated, can be made reasonably tractable if, as we shall see below, the Cartesian basis is used as well.   

The general (spatially non-flat) FRW line element is expressed in the Cartesian basis as
\be \label{FRW}
ds^2 =\, -\, dt^2 +\, a^2(t)\, \c_{ij}(x)\, dx^i dx^j \,,
\ee 
with 
\be \label{3FRW}
\c_{ij}(x) = \le(1 + \fr{k x^2} 4\ri)^{-2}\! \d_{ij} \,\,, \quad \mbox{where $\,\, x = \sq{\d_{ij}\, x^i x^j}$} \,.
\ee 
Therefore, while resorting to the BEC wave-function\,(\ref{WF1}), 
%
%
it is to be kept in mind that the variable $r$, or the proper radial coordinate distance, now has not only an implicit time-dependence, but also a non-linear dependence on the comoving distance $x$ as well:
\be \label{propdist}
r(x,t) =\, \sq{h_{ij}(x,t)\, x^i x^j} =\, x \le(1 + \fr{k x^2} 4\ri)^{-1} a(t) \,.
\ee
This obviously makes the computation of the quantum potential very tedious and complicated, in general. However, our task is relatively easier in this paper, as we focus on the study of the background level BEC cosmological evolution, for which it suffices to extract out only the purely time-dependent (or homogeneous) part of the general ($x,t$ dependent) quantum potential. Nevertheless, the computation still involves a large number of rigorous mathematical steps, upon carrying out which, and making certain algebraic simplifications, we obtain the following form of the `homogeneous' quantum potential:
\be \label{QPk}
\UQ (t) =\, \fr{3 \hbar^2}{m^2} \le[H(t) \, \dot{Y}(t) \,+\, \fr 4 {\s^2}
\le\{F(t) +\, \fr 2 {\s^2} +\, \fr{4k}{a^2(t)}\ri\}\ri] \,,
\ee
where
\bea
&& Y(t) =\, \fr{\ddot{R}(t)}{R(t)} +\, \fr{3 H(t) \,\dot{R}(t)}{R(t)} \,\,, 
\label{Wt} \\
&& F(t) =\, \dot{H}(t) +\, 5 H^2 (t) +\, \fr{2 H(t) \,\dot{R}(t)}{R(t)} \,\,,
\label{Ft}
\eea
with $R(t)$ and $\s$ being, respectively, the temporal modulating factor and the spread of Gaussian in the BEC wave-function\,(\ref{WF1}).

The last term of Eq.\,(\ref{QPk}) shows the spatially non-flat generalization of the expression for $\UQ(t)$ obtained in ref.
\cite{DSS-QB}
(see the Eq.\,(9) therein). In fact, the generalization is rather profound, as one may notice from the Friedmann equations, which, as illustrated earlier, stand as the cosmological evolution equations in the semi-classical picture as well, and which, for a non-zero $k$, read as:
\bea 
&& H^2(t) =\, \fr{\k^2} 3 \, \r(t) -\, \fr k {a^2(t)} \,\equiv\, \fr{\k^2} 3 \, \rt(t) \,, \label{cosm-eq1k}\\
&& \dot H(t) +\, H^2(t) =\, -\, \fr{\k^2} 6 \Big[\r(t) +\, 3 p(t)\Big] \equiv\, -\, \fr{\k^2} 6 \Big[\rt(t) +\, 3 \pt(t)\Big] \,, \label{cosm-eq2k}
\eea
where, for convenience, we have defined 
\bea 
&& \rt(t) :=\, \r(t) -\, \fr{3k}{\k^2 a^2(t)} \,,
\label{trho}\\
&& \pt(t) :=\, p(t) +\, \fr k {\k^2 a^2(t)} \,.
\label{tp}
\eea
Note that $\rt(t)$ is nothing but the critical density of the universe.

With the stipulated wave amplitude, $\, R(t) = \Rp a^{-3/2}(t) \,$ [{\it cf}. Eq.\,(\ref{Rt})], we therefore have
\bea
&& Y(t) =\, -\, \fr 3 2 \le[\dot{H}(t) +\, \fr 3 2 \, H^2(t)\ri] =\, \fr{3\k^2} 4 \, \pt(t) \,, \label{Wtk} \\
&& F(t) =\, \dot{H}(t) +\, 2 H^2 (t) =\, \fr{\k^2} 6 \Big[\rt(t) -\, 3\pt(t)\Big] \,. \label{Ftk}
\eea
Substituting these in Eq.\,(\ref{QPk}), and using Eq.\,(\ref{cosm-eq1k}), we obtain
\be \label{QP1k}
\UQ(t) =\, \fr{24 \hbar^2}{m^2 \s^4} \le[1 + \fr{\k^2 \s^2}{12} \le\{\Big[1 + \fr{3 \k^2 \s^2} 8 \, a(t) \, \pt\,'(t)\Big] \rt(t) - 3 \pt(t)\ri\} +\, \fr{2k\s^2}{a^2(t)}\ri] \,,
\ee
where the prime $\{'\}$ denotes $d/da$. Eq.\,(\ref{QP1k}) is the generalization of Eq.\,(\ref{QP1}) in the subsection \ref{sec:BEC-Cosm-eqs}. 

Now, the Raychaudhuri equation, or the second Friedmann equation (\ref{cosm-eq2k}), being independent of the value of $k$, we have the same expression (\ref{rhotot}) for the total energy density $\rho$, as in subsection \ref{sec:BEC-Cosm-eqs}, in terms of the total pressure $p$, the quantum potential $\UQ$, and the BEC and baryon densities, $\rBt$ and $\rbt$, given by Eqs.\,(\ref{rhoB}) and (\ref{rhob}) respectively. Since, by definition, $\rt + 3\pt = \r + 3p$, Eq.\,(\ref{rhotot}) can be recast as
\be \label{rhototk}
\rt\,(t) =\, \fr{\rbp +\, \rBp}{a^3(t)} -\, \fr 2 {\k^2} \, \UQ (t) -\, 3 \pt\,(t) \,.
\ee
Using this, we derive from Eq.\,(\ref{QP1k}) the following expression for $\UQ$ as a function of the scale factor $a$:
\be \label{QP2k}
\UQ(a) =\, \fr{6 \,\a^2}{1 +\, \a^2 \s^2 \,\ft(a)} \bigg[1 \,+\, \fr{\k^2 \s^2}{12} \bigg\{\le[\rbp +\, \rBp\ri] \fr{\ft(a)}{a^3} 
-\, 3 \le[1 + \ft(a)\ri] \pt(a)\bigg\} +\, \fr{2k\s^2}{a^2}\bigg] \,, 
\ee
upon denoting, as before, $\a = 2\hbar/(m \s^2)$ ~[{\it cf}. Eq.\,(\ref{alpha})], and defining
\be \label{fta}
\ft(a) :=\, 1 +\, \fr{3 \k^2 \s^2 \, a \,\pt\,'(a)} 8 \,.
\ee 
Eq.\,(\ref{QP2k}) is the generalization of Eq.\,(\ref{QP2}) obtained in subsection \ref{sec:BEC-Cosm-eqs}.

On the other hand, the equation\,(\ref{pU-eq}), which is derived in subsection \ref{sec:BEC-Cosm-eqs} from the conservation relation (\ref{consv-eq}) satisfied by $\r$ and $p$, holds for a non-vanishing $k$ as well. Remember that (\ref{pU-eq})
%
%
%
%
is the equation that determines the cosmic evolution profile, when solved simultaneously with that for $\UQ(a)$, which is the equation\,(\ref{QP2k}) now. Working out the solution is of course difficult enough, and even more so, compared to the $k=0$ case, as the high degree of non-linearity arising from the substitution of Eq.\,(\ref{QP2k}) in Eq.\,(\ref{pU-eq}) gets compounded with the technical complication posed by the last term in the former. In particular, a straightforward inspection reveals that, unlike in the $k=0$ case, no specific ansatz (such as $p' = 0$) will lead to an exact $k\L$CDM solution in the generalized picture. 

Nevertheless, we can consider taking the approach as before, that is to re-express Eq.\,(\ref{QP2k}) in terms of the parameters $\L$ and $\vep$, defined respectively by Eqs.\,(\ref{Lambda}) and (\ref{eps}) in subsection \ref{sec:BEC-LCDM}, and have the following generalization of Eq.(\ref{QP4}) derived in subsection \ref{sec:BEC-dynDE}: 
\be \label{QP4k}
\UQ(a) =\, \fr{\le(1 - 3\vep\ri) \k^2 }{1 - \vep \le[1 - \ft(a)\ri]} \bigg[\L +\, \fr \vep {2 \le(1 - 3\vep\ri)} \bigg\{\le[\rbp + \rBp\ri] \fr{\ft(a)}{a^3} 
-\, 3 \big[1 + \ft(a)\big] \pt(a) +\, \fr{12 k}{\k^2 a^2}\bigg\}\bigg] \,,
\ee
with 
\be \label{ft}
\ft (a) =\, 1 +\, \fr{9 \vep \, a \, \pt\,'(a)}{4 \le(1 -\, 3\vep\ri) \L} 
\ee
being the function given by Eq.\,(\ref{fta}), re-written accordingly. The purpose is to exploit the smallness of $\vep$, by expanding $\UQ(a)$ in powers of the latter and work out the solutions for the total energy density and pressure, $\r(a)$ and $p(a)$, either iteratively or perturbatively (in which case, $\r(a)$ and $p(a)$ are also expanded in powers of $\vep$). After all, the smallness of $\vep$ is beyond question in the limiting $\L$CDM case in subsections \ref{sec:BEC-LCDM} and \ref{sec:BEC-QB}, as it follows from a few merely physical considerations therein, such as that the BEC density must not exceed the total matter density of the universe.
Although this relies on the presumption $k=0$, one cannot expect a cosmological solution with a small $\vep$ to be affected much by a non-zero $k$ either, since wherever $k$ appears in the generalized expression (\ref{QP4k}) for $\UQ(a)$, in the last term or in the terms involving $\ft$ and $\pt$, it always carries a multiplicative factor of $\vep$. Moreover, a non-vanishing spatial curvature only implies $|k| = 1$ and not anything large enough to affect the order of magnitude of $\vep$ which multiplies it in every $k$-involving term in Eq.\,(\ref{QP4k}) for $\UQ(a)$ that determines the evolution profile of the universe. 

An expansion of the right hand side of Eq.\,(\ref{QP4k}) in powers of $\vep$ leads to
\be \label{QP-seriesk}
\UQ(a) =\, \k^2 \le[\L +\, \vep \le\{\fr{\rbp +\, \rBp}{2 a^3} \,-\, 3 \Big[\pt (a) +\, \L\Big] +\, \fr{12 k}{\k^2 a^2}\ri\} +\, \cO \big(\vep^2\big)\ri] \,,
\ee
whence, keeping terms up to $\cO \big(\vep\big)$, and using the definition\,(\ref{tp}) of $\pt(a)$, we solve Eq.\,(\ref{pU-eq}) to obtain
\be \label{p-approxk}
p (a) \,\approx\, \pp a^{-2s} -\, \L \Big(1 - a^{-2s}\Big) +\, \fr{3k}{\k^2 a^2} \Big(1 - a^{2\vep}\Big) \,,
\ee 
where $\pp$ is the value of $p$ at the present epoch, and $s \approx 1 - \vep \,$ [{\it cf}. Eq.\,(\ref{s})]. Substituting Eq.\,(\ref{p-approxk}) back in Eq.\,(\ref{QP-seriesk}), we get
\be \label{QP-approxk}
\UQ(a) \,\approx\, \k^2 \le[\L +\, \vep \le\{\fr{\rbp + \rBp}{2 a^3} \,-\, \fr{3 \le(\pp + \L - 3\k^{-2} k\ri)}{a^{2s}}\ri\}\ri] \,,
\ee
which when plugged, alongwith Eq.\,(\ref{p-approxk}), in Eq.\,(\ref{rhototk}), yields
\be \label{rho-approxk}
\r\,(a) \,\approx\, \fr \rmp {a^3} +\, \L \,-\, \fr{3 \le(1 - 2\vep\ri) \le(\pp + \L\ri)}{a^{2s}} -\, \fr{9k}{\k^2 a^2} \Big[1 - \le(1 - 2\vep\ri) a^{2\vep}\Big] \,, 
\ee 
where $\rmp = \le(1 - \vep\ri) \le(\rbp + \rBp\ri)\,$ [{\it cf}. Eq.\,(\ref{LCDM-rm0})] is the total matter density at the present epoch.

Consequently, eliminating the parameter $\pp$ via the constraint arising by virtue of setting $H = \Hp$ at the present epoch ($a = 1$), we obtain from the Friedmann equation\,(\ref{cosm-eq1k}),
\be \label{ratHub-1k}
\le[\fr{H(a)}{\Hp}\ri]^2 \,\approx\, \fr \Omp {a^3} +\, \OLp + \fr{1 - \Omp - \OLp}{a^{2s}} +\, 4 \le(1 - a^{2\vep}\ri) \fr{\Okp}{a^2} \,,
\ee
where $\Omp$, $\OLp$ and $\Okp$ respectively denote the matter density parameter, the $\L$-density parameter and the $k$-density parameter at $a = 1$.
Note that, while the equation\,(\ref{ratHub-1}) obtained for $k = 0$ in subsection \ref{sec:BEC-dynDE} gets modified by the last term in Eq.\,(\ref{ratHub-1k}), the factor $\le(1 - a^{2\vep}\ri)$ in the latter is quite small, for a small $\vep$, at all epochs, i.e. for all finite values of the scale factor $a$. This is regardless of the value of $\Okp$, which is a free parameter here, and not constrained to be equal to $\le(1 - \Omp - \OLp\ri)$, as in the standard $k\L$CDM scenario. However, since $\Okp$ cannot exceed unity (as otherwise the total energy density of the universe is negative) we can safely infer that the last term in Eq.\,(\ref{ratHub-1k}) is quite insignificant compared to the other terms therein, which constitute Eq.\,(\ref{ratHub-1}) and for which the parameter $\vep$ is already estimated to be very small (albeit not with a small enough $1\s$ margin that can highly enhance the BEC mass $m$ from its Hubble value $m_{_H}$). In other words, the near-$k\L$CDM solution obtained in subsection \ref{sec:BEC-dynDE}, albeit for $k=0$, prevails in presence of a non-zero $k$ as well. Hence, taking the argument around, the parametric estimations done in the preceding section still stand, and so do the outcome that the BEC mass is comparable to that of a scalar field DE, rather than that of an axion-like DM. 

\section{Conclusion} \label{sec:concl}

We have thus explored the possible ways of realizing a semi-classical picture of a {\em unified} cosmic dark sector, from a BEC of ultralight bosons stretching across a very large length scale, close to the Hubble scale. 
While this implies a {\em quantum} `dark fluid' description of the universe being sought from the BEC, our primary objective in this paper has been a self-consistent theoretical formulation, from the foundations laid in some of our earlier works
\cite{db1,db2,db3,DSS-QB}. 
In particular, we have pondered on the essential criteria for realistic scenarios in the standard paradigm of the spatially flat FRW cosmology, such as the conditions for the formation of the BEC in the early universe and the spherical symmetry of the corresponding wave-function $\Psi$, apart from suggesting the functional form of the latter, so that the BEC may source the CDM content of the universe, albeit having interactions with the effective DE component in a unified picture. 

The key role in such a unification is shown to be played by the associated quantum potential $\UQ$, derived from the amplitude of $\Psi$, as the bosons in the BEC being (supposedly) quantum particles, follow the (Bohmian) quantal trajectories
\cite{bohm-QT1,bohm-QT2,BHK-QT}.
In fact, it is this quantum potential $\UQ$ which leads to a quantum correction to the Raychaudhuri-Friedmann equation
\cite{sd},
and thereby makes an effective DE component perceptible in the formalism. While this has been pointed out in the earlier papers
\cite{db1,db2,db3,DSS-QB},
quite in contrast with a variety of BEC-DM models in the literature
\cite{HBG-FDM,wang-BEC_DM,FM-BEC_DM,FMT-BEC_DM,BH-BEC_DM,lopez-BEC_DM,sik-BEC_DM,chav-BEC_DM,harko-BEC_DM,DG-BEC_DM,KL-BEC_DM,HM-BEC_DM,SRM-BEC_DM,BCL-BEC_DM,LRS-BEC_DM,li-BEC_DM,EMM-BEC_DM,DKG-BEC_DM,giel-BEC_DM,david-BEC_DM,DKKG-BEC_DM,SCB-BEC_DM},
what we have demonstrated in this work is the natural emergence of the unified cosmic dark sector, from the perception of the probability density $|\Psi|^2$ as the energy density $\rBt$ of a dust-like fluid component of the universe, semi-classically. Note that the approach is semi-classical also in the sense that the density $\rbt$ of the visible matter content of the universe, which complements the BEC density $\rBt$ and the quantum potential $\UQ$, is taken to be of the form of that of the standard non-relativisitic fluid, viz. the baryonic dust. Moreover, the entire formalism here concerns the cosmological evolution at the background (or homogeneous) level, with no allusion to the perturbative spectrum whatsoever. Nor we have looked into the detailed aspects of the BEC, i.e. about the types of its constituent bosons, and interactions thereof.

Now, to physically realize the dark universe picture, i.e. to examine the viability of the same, one requires to work out in the first place, the suitable solution(s) of the BEC cosmological equations. While this seems to be a formidable task, since $\UQ$ turns out to be a highly non-linear function of the energy density and pressure of the system, $\rho$ and $p$ (and derivatives thereof), a simple ansatz of the constancy of $p$ at once leads to a solution describing an effective $\L$CDM cosmic evolution. However, neither the BEC density $\rBt$ accounts for the effective CDM density in entirety, nor the quantum potential $\UQ$ is solely responsible for the the effective cosmological constant $\L$. Instead, the entire bulk of physical constituents of the universe (including even the visible baryons) back-reacts on the metric structure of space-time. Such a {\em quantum back-reaction} (QB) effect is quite intriguing and significant in the sense that the positive-definite and dimensionless parameter $\vep$, which determines its strength, roughly varies as the inverse of the BEC mass $m$. As such, the lower bound on $\vep$ determines the extent to which $m$ exceeds its Hubble value $m_{_H} \simeq 10^{-32}\,$eV. 

Progressively tighter constraints on $\vep$ are shown to follow from certain purely physical considerations, such as the positive-definiteness of the total matter density of the universe at the present epoch, $\rmp$, the positive-definiteness of the BEC mass squared, and the stringent demand that $\rmp \geq \rBp$, where $\rBp$ is the value of the BEC density $\rBt$ at the present epoch. However, even the tightest constraint thus found, which implies $\vep$ is $\cO(0.1)$ or less, is loose enough for the corresponding mass ratio $m/m_{_H}$. Specifically, one cannot conclusively say anything when it comes to comparing this constraint with the stringent observational mass limits (typically $\gtrsim 10^{-24}\,$eV) obtained under the demand of suppressing small scale structures of commonly known ultralight bosonic dark matter candidates, such as {\em axions} 
\cite{MH-UBDM,fer-UBDM,marsh-UBDM,MN-UBDM,APYMB-UBDM,CAMD-UBDM,HOTW-UBDM,HGMF-UBDM}.
This makes it imperative to estimate $\vep$ directly and robustly using the observational data, and that too for a solution of the BEC cosmological equations as general as possible. At least, a solution more general than $\L$CDM needs to be worked out, which should explicitly have $\vep$ appearing in the expression for the Hubble parameter $H(z)$, in order that a statistical estimation of $\vep$ could be done by using the commonly considered SN-Ia Pantheon data and the observational Hubble data. 
\cite{scol,panth,MPCJM-OHD,YRW-OHD,RDR-OHD,mors}.

While looking for such a general solution, one is nonetheless compelled to confront with the analytical complication posed by complicated form of the quantum potential $\UQ$. However, we have had at least the smallness of $\vep$ to some avail. Therefore, expanding $\UQ$ in powers of $\vep$, and retaining terms up to the linear order, we have obtained a solution more general than $\L$CDM and with the corresponding expression for $H(z)$ inevitably involving $\vep$. The observational constraints on $\vep$ and other relevant parameters, such as the total matter density parameter at the present epoch, $\Omp$, and the Hubble constant $\Hp$, are consequently worked out following the standard $\chi^2$ minimization technique, and using the Metropolis-Hastings algorithm for MCMC. Apart from a marginal deviation from $\L$CDM, the estimates show a BEC mass enhancement (from the Hubble value $m_{_H} \simeq 10^{-32}\,$eV) by about three orders of magnitude, up to $1\s$. 
%
So, the BEC mass is comparable to that of a scalar field DE, which is of course contrary to the usual perception of the typical DM candidature of the BEC. However, this is an outcome that stands even when the assumption of the spatial flatness of the universe is dropped, which we have shown by carrying out a rigorous generalization of the BEC cosmological equations and finding the corresponding solution by exploiting the smallness of the quantum back-reaction parameter $\vep$ once again.     



\section*{Acknowledgment}

This work has been supported by the Natural Sciences and Engineering Research Council of Canada.
SS acknowledges financial support from Faculty Research Programme Grant -- IoE, University of Delhi (Ref.\,No./IoE/2023-24/12/FRP).


\section*{Data Availability Statement}

The manuscript has no associated data or the data will not be
deposited. Observational data-sets used for the statistical estimation are
retrieved from well-known references, duly cited in this paper.

\bigskip


\end{document}